\documentclass[referee,useAMS,usenatbib,usegraphicx]{mn2e}
\usepackage{natbib}
\usepackage{amssymb}
\usepackage[usenames, dvipsnames]{color}
\title[Low-mass young stellar population of the cluster IC~1805]{Low-mass young stellar population and star formation history of the cluster IC~1805 in the W4 H{\sc ii} egion}
\author[Panwar et al.]
{Neelam Panwar$^{1,2}$\thanks{E-mail:neelam$\_$1110@yahoo.co.in}, M. R. Samal$^{2,3}$,  A. K. Pandey${^4}$, J. Jose${^5}$, W. P. Chen${^2}$, 
\newauthor D. K. Ojha${^6}$, K. Ogura${^7}$, H. P. Singh${^1}$, R. K. Yadav${^8}$\\
$^1$Department of Physics \& Astrophysics, University of Delhi, Delhi - 110007, India\\
$^2$Graduate Institute of Astronomy, National Central University 300, Jhongli City, Taoyuan County - 32001, Taiwan\\
$^3$Laboratoire d'Astrophysique de Marseille-LAM, Universit\'e d'Aix-Marseille \& CNRS, UMR7326 13388 Marseille CEDEX 13 France\\
$^4$Aryabhatta Research Institute of Observational Sciences (ARIES), Nainital - 263129, India\\
$^5$Kavli Institute for Astronomy and Astrophysics, Peking University, 5 Yiheyuan Road, Haidian District, Beijing 100871, P. R. China\\
$^6$Tata Institute of Fundamental Research, Mumbai (Bombay) - 400 005, India\\
$^7$Kokugakuin University, Higashi, Shibuya-ku, Tokyo - 1508440, Japan\\
$^8$National Astronomical Research Institute of Thailand (NARIT), 50200, Thailand}
\begin{document}

\date{}
\pubyear{2017}

\maketitle

\label{firstpage}

\begin{abstract}

W4 is a giant H{\sc ii} region ionized by the OB stars of the cluster IC~1805.
 The H{\sc ii} region/cluster complex has been a subject of numerous investigations 
as it is an excellent laboratory for studying the feedback effect of massive stars on the surrounding 
region. However, the low-mass stellar content of the cluster IC~1805 remains poorly studied till now. 
With the aim to unravel the low-mass stellar population of the cluster, we present the results 
of a multiwavelength study based on deep optical data obtained with the Canada-France-Hawaii Telescope, infrared data from 
2MASS, $Spitzer$ Space Telescope and X-ray data from $Chandra$ Space Telescope. The present optical dataset is complete enough 
to detect stars down to 0.2~M$_\odot$, which is the deepest optical observations so far for the cluster. 
We identified 384 candidate young stellar objects (YSOs; 101 Class I/II and 283 Class III) within the cluster using various 
colour-colour and colour-magnitude diagrams. We inferred the mean age of the identified YSOs to be $\sim$ 2.5 Myr and 
mass in the range 0.3 -- 2.5 M$_\odot$. The mass function of 
our YSO sample has a power law index of -1.23 $\pm$ 0.23, close to the Salpeter value (-1.35), 
and consistent with those of other star-forming 
complexes. We explored the disk evolution of the cluster members and found that the diskless sources are relatively older compared to the disk bearing YSO candidates. 
We examined the effect of high-mass stars on the circumstellar disks and found that within uncertainties, the influence of massive stars 
on the disk fraction seems to be insignificant. We also studied the spatial correlation of the YSOs with the distribution of gas and dust of the complex to conclude that IC 1805 would have formed in 
a large filamentary cloud.

\end{abstract}
\section{Introduction}
 Most recent studies show that stars are formed in clusters or groups and nearly half of the 
low-mass populations are born in massive young clusters or OB associations \citep{lada03,all07}.
 Young clusters are natural laboratories for testing the star formation 
processes and the stellar evolutionary theory as the cluster members have a wide mass spectrum 
while sharing the same chemical composition and distance.

However, the presence of very luminous (O- or early B-type) stars in such systems profoundly influences their environment by compressing 
and sweeping the surrounding material because of their strong ionizing radiation and powerful 
stellar winds \citep[see e.g.,][]{pre02,zav07,pov09,cha09,cha11,deh10,jos13,sam14}. Ultraviolet (UV) radiation from massive stars not only
 affects the surrounding molecular cloud and its star formation but also causes photoevaporation of disks around the 
young low-mass stars in their vicinity \citep[see e.g.,][]{bally98,adam04,bal07,gorti09}, thereby playing a key role in shaping the fundamental properties of 
the cluster such as stellar mass function (MF), total 
star formation efficiency, and evolution of circumstellar disks around the young stars.   

IC~1805, also known as Melotte~15 or OCL~352, is a young open cluster associated with the H{\sc ii}
 region W4 (the Heart Nebula) which is a part of the W3-W4-W5 complex, and contains dozens of massive 
 OB-type stars \citep{mas95}. The W4 complex evacuated by the combined energetic winds of the OB stars within IC~1805
\citep{gou80,nin95}, appears to be a void in the atomic hydrogen layer with a width of about 100~pc \citep{tay99}. The W4 complex also contains several cometary-shaped bright-rimmed clouds (BRCs) or pillars, pointing towards the luminous stars 
of IC~1805 \citep{sug91,lef97}, similar to those found in \citet{ogu02}, and \citet{cha11}. Such structures are generally the result of 
compression and erosion of pre-existing clouds by massive stars \citep[][]{bis09,mia09}, indicating  
that stellar feedback is playing a strong role in altering the morphology and physical conditions
 of the W4 complex. Therefore, the complex has been a subject of numerous investigations for understanding 
the role of massive stars on the star formation process in the surrounding cloud
 \citep[e.g.,][]{lad78,tay99,oey05,koe12,jose16}.

IC~1805 is located at a moderate distance of $\sim$ 2.0 kpc \citep{str13}, similar to the adjacent H{\sc ii} regions W3 \citep[2.0 kpc,][]{hac06,xu06} and W5~E \citep[2.1 kpc,][]{cha11}. 
It is embedded in a very low extinction cloud of $E$($B - V$) $\sim$ 0.8 mag or visual extinction (A$_V$) $\sim$ 2.4 mag \citep{joshi83,str13}. 
Also, according to \citet{gue89}, \citet{sagr90}, and \citet{han93} the extinction law 
in the direction of IC 1805 is normal. Hence, it is a good target for studies on star formation and evolution of 
low-mass stars; in particular, how they are influenced by OB type stars.  
Though the cluster has been investigated by several authors at optical bands 
\citep{joshi83,kwo83,sag87,gut89,mas95,sung95,nin95}, most of these studies were limited to the high-mass stars only.  
These observations were not 
only too shallow to detect the low-mass stellar populations but also cover 
only a relatively small area of the complex. Though several attempts have been made to identify young stellar objects (YSOs) 
in the complex \citep{koe12,pov13,str13,bro13}, 
no adequate attention has been paid to the characterization of low-mass young stellar population 
of the cluster using deep optical photometric data. 

 Since low-mass stars constitute the majority of the stellar population in a cluster and even in the Galaxy \citep{kro02}, 
the low-mass 
stellar content and its MF are essential to understand the 
nature of the star formation process and the properties of open clusters. Young stars can be identified using multiwavelength observations depending on 
their evolutionary stages and  
spectral types \citep[e.g.][]{car06,dah07}. Overall, the
$Spitzer$ results suggest that though primordial circumstellar disks around 
solar and sub-solar mass stars can last for up to 10 Myr, disk lifetimes are a factor of 2 shorter for higher mass objects 
\citep[see][references therein]{wil11}. 

The infrared (IR) observations are particularly useful to identify pre-main-sequence (PMS) stars exhibiting IR excess emission. %In the case of PMS 
Also, with the help of X-ray data, young stars can be identified from their excess X-ray emission  
due to magnetic reconnection flares similar to those seen on the solar surface, but with X-ray 
fluxes 2 to 3 orders of magnitude larger in comparison to solar-type stars \citep{fei99,pre05,gue07}. 
While near infrared (NIR), mid-infrared (MIR) and X-ray observations
 are suitable to identify PMS stars, they have limitations in characterization of these stars 
themselves (e.g., age and mass determinations). This can be done best by optical 
observations as PMS stars possess little or no circumstellar emission at optical wavelengths. 

In the present work, we attempt to investigate the low-mass
 stellar content of the cluster IC~1805 based on high spatial resolution, deep optical photometric 
observations taken with the Canada-France-Hawaii Telescope (CHFT) 
in combination with $Spitzer$, Two 
Micron All Sky Survey (2MASS), and X-ray data available for the region. 
We identify and characterize a sample of low-mass young cluster members with circumstellar disks
 and evolved diskless
 candidate members. %which has been identified by using Spitzer-IRAC/MIPS, Chandra X-ray data,  
We examine the physical properties of the cluster (e.g., cluster extent, age, disk evolution and mass-function) 
 and try to understand the star formation process in this region. 
\section{Observations and Data Reduction}
\begin{figure}
\centering
\includegraphics[scale = 0.5, trim = 0 0 0 0, clip]{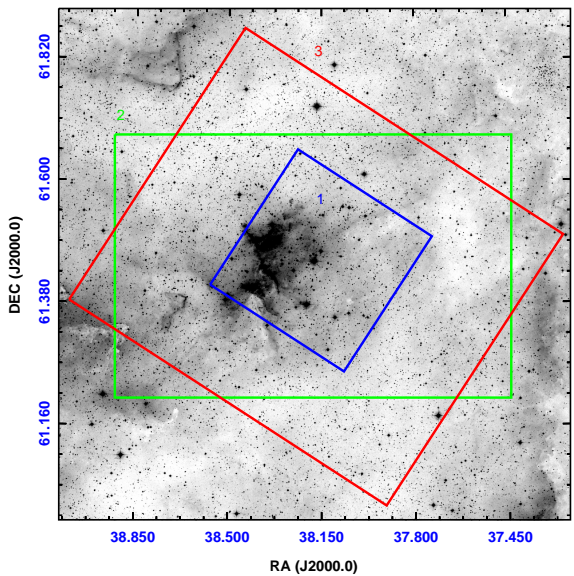}
\caption{The DSS2-R image of the IC~1805 H{\sc ii} region. The area covered by X-ray, optical and $Spitzer$-IRAC observations are 
shown with blue, green and red colours, respectively (also marked as `1', `2' and `3'). }
\label{fig1}
\end{figure}
Fig. \ref{fig1} shows the DSS2-R image of the IC~1805 region. The regions marked in the blue, green and red are the areas 
% `1', `2' and `3' represent the area 
covered by X-ray, optical and $Spitzer$-Infrared Array Camera (IRAC) observations, respectively. Multi-band Imaging photometer for 
{\it Spitzer} (MIPS) observations are available for the entire area covered by the IRAC observations. The individual observations and 
photometric catalogs are described below.
\subsection{Optical Observations}
The pre-processed deep $V$ and $R$ band images of the IC~1805 region were obtained from the CFHT archive. The data were obtained 
with CFH12K, a mosaic CCD camera on CFHT, on 2002, January 07 (P. I. - Sung H.). With a plate scale of 
0.21 arcsec per pixel, CFH12K provides a 42$^\prime$ $\times$ 28$^\prime$ field of view (FOV). Three exposures of 150s were taken in $V$ and $R$ 
bands each and the typical seeing during the observations was $\sim$ 0$^{\prime\prime}$.9.
 We obtained the photometry of the stars 
by using the standard $IRAF$ tasks and $DAOPHOT-II$ 
software package \citep{ste87}. We used the DAOFIND task in $IRAF$ to extract
the point sources, and selected only those sources
having S/N 5$\sigma$ above the background. We then performed point-spread function (PSF) photometry on the selected sources using the ALLSTAR routine. 
 These deep optical observations were obtained from the archive
with the aim to characterize the low-mass members (in particular to obtain  the age and mass of the 
low-mass young stellar population) of the cluster IC 1805. To calibrate these deep images, we obtained the {\it V} and {\it R}
 band observations of the central region of IC~1805 on 2012, October 26 
by using a 2048 $\times$ 2048 pixel$^2$ CCD camera mounted at the {f/4} Cassegrain focus of 
the 1.3-m Devasthal Optical Telescope (DOT) at Aryabhatta Research Institute of Observational 
Sciences (ARIES), Nainital, India. The CCD camera 
with a pixel size 13.5 $\micron$ and a plate scale 0.54 arcsec per pixel, covers a $\sim$ 18$^\prime$ $\times$ 18$^\prime$ FOV. During the 
observations the seeing was $\sim$ $1^{\prime\prime}.5$. To standardize the observations, the SA~98 field of \citet{lan92} 
was also observed on the same night at various airmasses. A number of bias and twilight flat frames 
were also taken during the observations.

The DOT images were reduced using various tasks available in the $IRAF$ software package. Instrumental magnitudes were 
obtained by using the $IRAF/DAOPHOT-II$ package via PSF fitting. The PSF was obtained for each 
frame by using several isolated stars in the frames. The atmospheric extinction coefficients and zero-point magnitudes obtained from 
the SA~98 standard field observations are given in Table 1. The instrumental magnitudes were
 converted to the standard values by using 
a least-square linear regression procedure outlined by \citet{ste92}. 

\begin{table}
\caption{The zero-point constants, colour coefficients and extinction coefficients}
%begin{minipage}{15mm}
%caption{label{Log of observations}}
\begin{tabular}{|p{1.2in}|p{1.2in}|}
\hline
Parameters &  Constants\\
\hline
Zero-point constants &  \\
c1 &-0.226 $\pm$ 0.016  \\
c2 &3.114 $\pm$ 0.014\\  
Colour coefficients &  \\
m1 &1.058 $\pm$ 0.022   \\
m2 &0.099 $\pm$ 0.024    \\  
Extinction coefficients&  \\
K$_v$&0.158 $\pm$ 0.015  \\
K$_r$&0.109 $\pm$ 0.011  \\
\hline
\end{tabular}
\end{table}
The following transformation
equations were used to calibrate the observations:

{\it $(V-R) = m_1 (v - r) + c_1 $,}

\hspace{0.4cm}{\it     $V= v + m_2 (v - r) + c_2 $,}

where {\it v, r} are the instrumental magnitudes corrected for the atmospheric
extinctions, and {\it V, R} are the standard magnitudes; {\it c$_1$, c$_2$}
and {\it m$_1$, m$_2$} are zero-point constants and colour coefficients,
respectively. 
The final photometry from the DOT observations is used to calibrate the CFHT photometry. 
We obtained 27,427 sources with a magnitude uncertainty $\le$ 0.2 mag in both $V$ and $R$ bands.

\subsection{$SPITZER$ IRAC Observations} 

The NIR/MIR data from the {\it Spitzer} space telescope \citep{wer04} using IRAC 
\citep{faz04} centered at 3.6, 4.5, 5.8 and 8.0 $\micron$, were obtained from the 
$Spitzer$ archive\footnote{http://archive.spitzer.caltech.edu/}. 
The IRAC camera provides simultaneous 5$\arcmin$.2 $\times$ 5$\arcmin$.2 images in four bands 
at 3.6, 4.5, 5.8 and 8.0 $\micron$ and  at a spatial resolution of $\sim$ 2 arcsec.
IC~1805 was observed in 2006, September (P.I. - S. Wolff, Program ID - 20052). 
The corrected Basic Calibrated Data (cBCD) images 
of the region ({\it Spitzer Science 
Center}'s IRAC pipeline, version S18.14.0) were obtained by using the Leopard software. 
The images were taken in High Dynamical Range 
mode. Both short 
(0.6s) and long (12s) integration cBCD frames in each channel were
 separately mosaicked using $MOPEX$  (version 18.3.1). 
The $MOPEX$ - $APEX$ pipeline was used to detect the point sources and to perform the point response function (PRF) 
fitting photometry. Since the nebular emissions can mimic point-like sources, especially in 5.8 and 8.0 $\micron$, 
all the sources were visually examined and ambiguous sources were removed. In addition, we also included point sources manually which were 
not automatically detected by $APEX$ and supplied the list of these sources to $Apex-user-list$ pipeline to  
perform the PRF photometry. We have adopted the zero-points for the conversion between 
flux densities and magnitudes to be 280.9, 179.7, 115.0 
and 64.1 Jy in the 3.6, 4.5, 5.8 and 8.0 $\micron$ bands, respectively, as given in the IRAC Data Handbook 
(Reach et al. 2006). The sources with photometric uncertainties 
$\leq$ 0.2 mag in each band were considered as good detection. 
To obtain the final catalogue of the sources detected in all IRAC bands, we made a catalog for each channel 
from the short and long exposures separately 
and then looked for the closest match within 1$^{\prime\prime}$.2. The final catalog contains 32,014 sources which are detected at least 
in two IRAC bands, of which 5288 sources have photometry in all IRAC bands. 

\subsection{$SPITZER$ MIPS Observations} 

The region was observed at MIPS 24 $\micron$ in 2005 September (P.I. - J. S. Greaves, Program ID - 3234). %BRC 5 \& 7, were not fully covered in these observations. 
We downloaded the MIPS post-BCD images from the archive, which were created at the image scale of 
 2$^{\prime\prime}$.45 per pixel and at a spatial resolution of $\sim$ 6$^{\prime\prime}$ . We supplied a list of coordinates of visually identified sources to the $Apex
-user-list$ module and performed the PRF fitting to extract the fluxes of the sources. The zero point 
value 7.14 Jy (adopted from MIPS Data Handbook\footnote{http://irsa.ipac.caltech.edu/data/SPITZER/docs/mips/mipsinstrumenthandbook/}) is used to convert the flux densities to magnitudes. We cross matched the MIPS catalogue to the IRAC source catalogue using a matching radius of 2$^{\prime\prime}$.5 \citep{meg12,jose16} and found that 
 164 sources have counterparts in one or more IRAC bands.

\subsection{Near-infrared photometry from 2MASS}
NIR $JHK_s$ data for the point sources within the H{\sc ii} region (shown in Fig. 1) have been obtained from the 2MASS 
Point Source Catalog \citep[PSC,][]{cut03}. Sources 
with uncertainty $\leq$ 0.2 mag and quality flag `AAA' in all the three bands were selected 
to ensure good quality data. 
The spatial resolution of 2MASS JH$K_s$ 
bands  is $\sim$ 2$^{\prime\prime}$.
\section{Results}
\subsection{Identification and Classification of Young Stellar Objects}
Full understanding of cluster properties and its star formation history requires proper identification
 of cluster members. In the absence of spectroscopic observations, one robust approach is to identify young stellar 
 content using multiwavelength photometric observations. 
Since IC~1805 is located at a low Galactic latitude, the contamination from background/foreground 
stars may significantly affect the analyses at infrared bands 
\citep[e.g.][]{get06,roc11}.  Hence, careful identification and classification of probable YSO members 
is essential. During their early phase, stars possess cirmcumstellar disks and hence show excess emission at IR 
wavelengths. The IR excess properties of YSOs can be used to identify and classify them. 
Class `0' 
sources emit most of their radiation  
in far infra-red (FIR) to submillimetre regime of the electromagnetic spectrum \citep{and95}. In the Class~`I' phase, YSOs 
 radiate mostly in MIR-FIR wavelengths, whereas Class~`II' sources 
exhibit NIR-FIR excess emission. Class~`III' sources 
possess little or no excess in IR \citep{lada87,ada87,and93,and95} but show enhanced 
X-ray activities as explained in Section 3.1.4. However, sources which lack inner disks and show 
excess emission in MIR wavelengths are referred to 
as `transition disk' sources \citep{muz06,her06}. We used optical, $Spitzer$-IRAC-MIPS, 2MASS and X-ray datasets to identify and classify the 
YSOs in IC~1805. The various selection criteria adopted for identification and classification
 of those YSOs are given below.    
\begin{figure}
\centering
\includegraphics[scale = 0.35, trim = 10 10 10 10, clip]{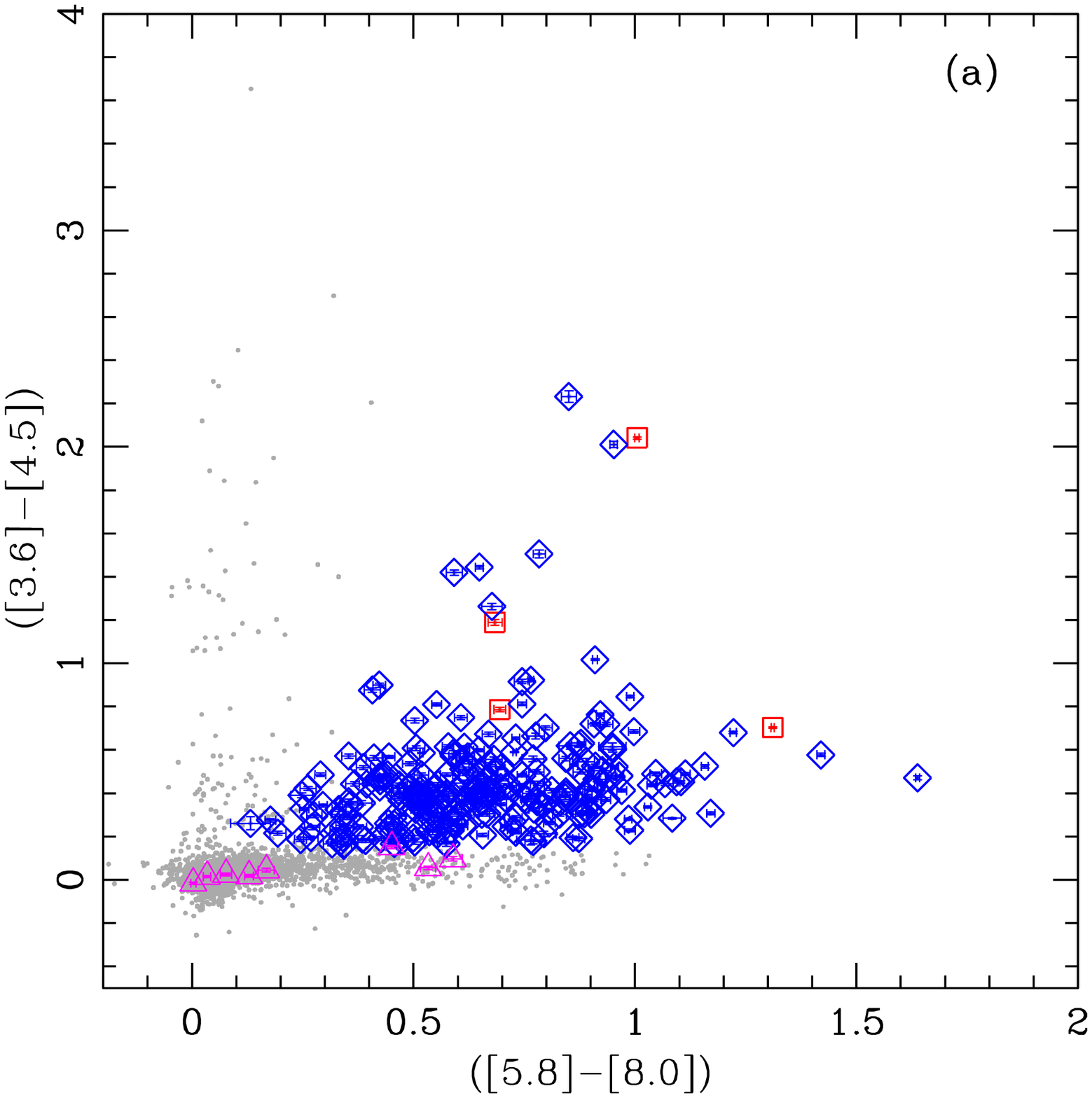}
\includegraphics[scale = 0.35, trim = 10 8 8 10, clip]{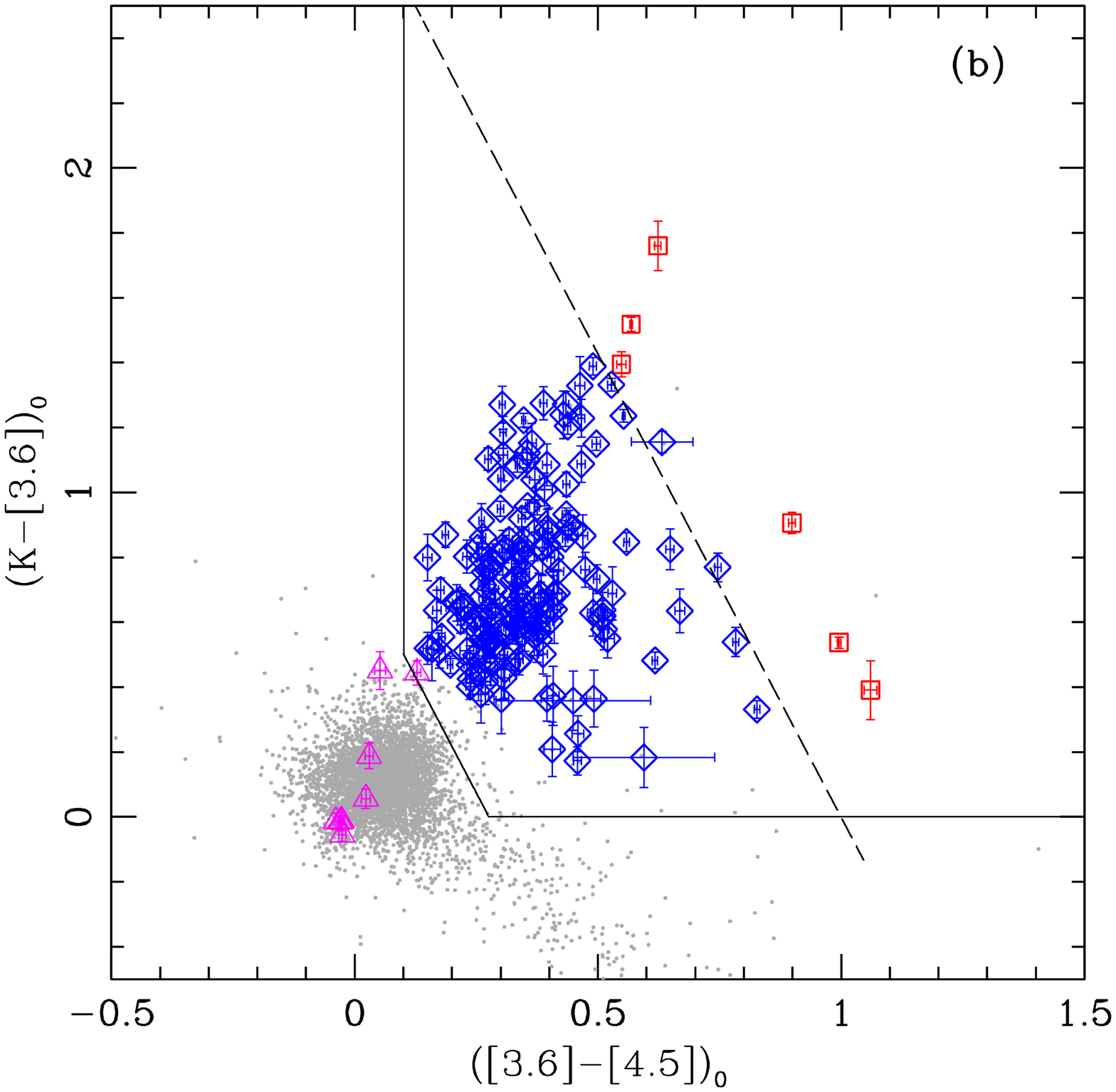}
\includegraphics[scale = 0.35, trim = 10 10 10 10, clip]{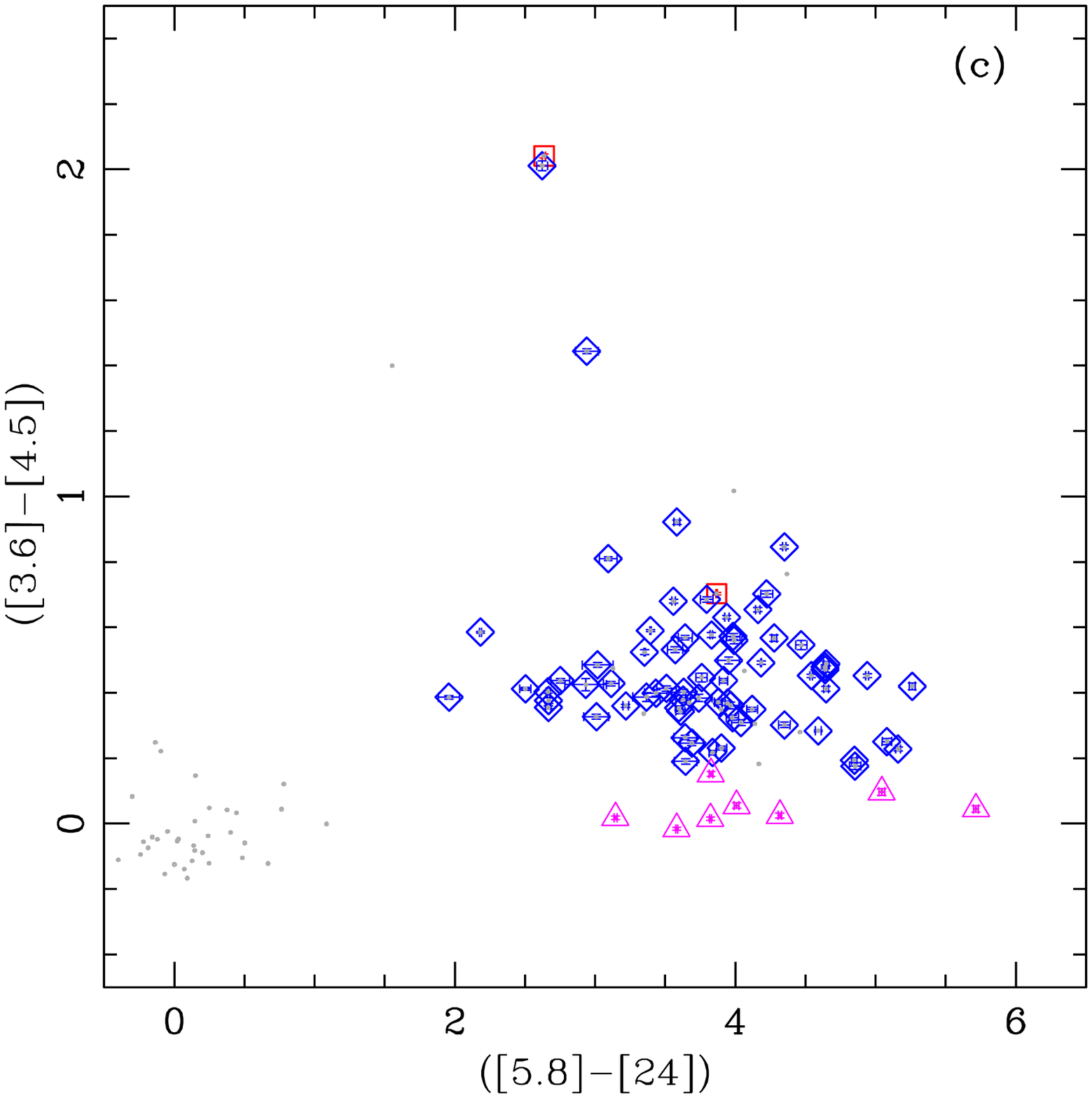}
\caption{ (a) IRAC ([3.6]-[4.5])/ ([5.8]-[8.0], (b) IRAC/2MASS, and (c) IRAC/MIPS colour-colour 
diagrams for the sources (black dots) in IC~1805. The error bars represent the 1-$\sigma$ uncertainty in respective colours. 
Class~I and Class~II sources are shown with red square and blue diamond symbols, respectively. Transition disk candidates are shown with magenta triangles.}
\label{fig2}
\end{figure}

\subsubsection{YSOs selected by IRAC colours}
Since YSOs occupy distinct positions on the IRAC colour-colour diagrams due to their different characteristics, 
the colour-colour diagrams are widely used to identify and classify YSO population in a star-forming region 
\citep{meg04,all04}. 
\citet{gut09} developed a method consisting of a series of IRAC/MIPS colour criteria to identify 
and classify YSOs. Though the mid-IR and far-IR observations can penetrate deep into the thick layers of 
dust and gas and reveal the embedded YSO population, they also enhance the chances of inclusion 
of background contaminants 
in the YSO sample. For example, polycyclic aromatic hydrocarbon (PAH)-emitting galaxies, active 
galactic nuclei (AGN), unresolved blobs of shocked emission etc., may mimic 
YSOs and, thus, can contaminate our YSO sample. In order to weed out possible contaminants, 
we used the customized cuts in the $[4.5] - [5.8]$ vs. $[5.8] - [8.0]$ and $[3.6] - [5.8]$ vs. $[4.5] 
- [8.0]$ colour-colour spaces, as described by \citet{gut09}.  
After excluding the contaminants based on the \citet{gut09} approach, we identified 238 YSO candidates. 
 Of these, 4 sources satisfy the colour criteria of Class~I sources ([4.5] -- [5.8] $>$ 0.7 \& [3.6] -- [4.5] $>$ 0.7) 
and 234 sources (satisfying the following conditions) were classified as Class~II candidates : 

$[4.5] −- [8.0] - \sigma >$ 0.5,
 
$[3.6] −- [5.8] - \sigma$ $>$ 0.35, 

$[3.6] −- [5.8] + \sigma$ $\le$ $(0.14/0.04) \times (([4.5] −- [8.0] -− \sigma) -− 0.5) + 0.5$, and 

$[3.6] −- [4.5] - \sigma$ $>$ 0.15;

Here, $\sigma$ is the photometric uncertainty in respective colours. The rest of the sources may be 
Class~III or field stars.  
 However, it is to be noted that in a few cases, due to reddening and disk inclination effects, a Class~II source 
could mimic the colours of 
a Class~I source \citep{whi03,gut09}. Fig. \ref{fig2}a shows the $[3.6] −- [4.5]$ vs. $[5.8] -− [8.0]$ colour-colour 
diagram for all the uncontaminated IRAC sources. The Class~I and Class~II sources are shown as 
red squares and blue diamonds, respectively.  

\subsubsection{YSOs from 2MASS/IRAC} 
At the central region of W4, in the vicinity of IC 1805, the background emission at 5.8 and 8.0 $\micron$ 
is high compared to the emission at 3.6 and 4.5 $\micron$. This high background emission
 is often observed in massive star forming regions \citep[e.g.,][]{ind07,koe08,ojha11}. 
Many of the sources detected in 3.6 and 4.5 $\micron$ may be missing in 5.8 and 8.0 $\micron$ bands 
due to the enhanced nebulosity of the background. 
To ascertain the nature of those sources, we combined the IRAC 3.6 and 4.5 $\micron$ data with those of 2MASS H and K$_s$ bands. 
\citet{gut09} used the intrinsic ($K_s - [3.6]$) vs. ($[3.6] - [4.5]$) colour criteria to identify the sources which are detected 
in 3.6, 4.5 $\micron$ and possess counterparts in 2MASS bands. 

To identify the YSOs using IRAC and 2MASS data, we first created an extinction map 
based on 2MASS data following the method discussed in \citet{pan14}, and then derived the intrinsic colours 
of the YSOs by correcting for the extinction values read off from this map. 
 To generate the extinction map, we divided the region into 
small cells and computed the colour excess in each cell using the relation $E$ = ($H - K_s$)$_{obs}$ -- ($H - K_s$)$_{int}$,
 where ($H - K_s
$)$_{obs}$ is the observed 
median colour of the stars in a cell and $(H - K_s)_{int}$ is the intrinsic median colour of the control field stars. The colour excess ratios 
presented in \citet{fla07} have been adopted to calculate the extinction in $K_s$ band ($A_{K_s}$) 
for each cell by using the relation $A_{K_s}$=1.82 $E$. Based on the extinction map,
the mean $A_{K_s}$ in the cluster direction corresponds 
to $A_V$ $\sim$ 2.6 mag, which is comparable to the value reported in the literature \citep{joshi83}.
We calculated the $(K_s-[3.6])_0$ and $([3.6]-[4.5])_0$ colours of the sources  
and used various colour cuts as described in \cite{gut09} to select Class~I and Class~II sources.
By this method, we obtained 161 Class~II and 6 Class~I candidates. Among these, based 
on the IRAC colours, 5 Class~I sources have been classified as Class~II (Sec. 3.1.1), and for further analyses we have considered 
them as Class~II candidates. Out of these 167 Class I/II candidates 133 were identified by using IRAC colour-colour diagram.  
Fig. \ref{fig2}b shows the 2MASS/IRAC colour-colour diagram for sources having K$_s$, 3.6 and 4.5 $\micron$ detection. 

\subsubsection{YSOs from MIPS 24 $\micron$} 
To identify the transition disk candidates, we used the 24 $\micron$ data in addition to 4.5 and 5.8 $\micron$ 
data. Following \citet{muz06} and \citet{her06}, sources mainly with photospheric radiation 
in the IRAC or IRAC/2MASS classification but having excess emission at 24 $\micron$ 
($[5.8] - [24]$ $>$ 2.5 and/or $[4.5] - [24]$ $>$ 2.5) are reconsidered as transition disk sources. 
 We identified 8 such sources. Since
 heavily embedded Class II sources can be misclassified as Class I based on the IRAC colour criteria (see Section 3.1.1), 
those Class~I sources with 24 $\micron$ data were re-checked to ensure that their SEDs rise from the IRAC to MIPS bands. 
All Class I candidates were considered as Class~II, if they do not have $[5.8] - [24]$ $>$ 4 or $[4.5] - [24]$ $>$ 4. 
Out of 4 Class~I sources identified in Section 3.1.1, 2 are re-classified 
as Class~II, while the remaining 2 do not possess 24 $\micron$ measurements. 
Fig. \ref{fig2}c shows the IRAC/MIPS colour-colour diagram for the sources in the region. In Fig. 2, transition disk candidates based on the MIPS photometry are marked as magenta filled circles.
 
\subsubsection{YSOs from $Chandra$ X-ray Data}
YSOs are often strong X-ray emitters compared to their main-sequence 
(MS) counterparts \citep[e.g.,][]{fei99,get05,pre05,gue07}. 
Hence, X-ray surveys of star forming regions are used to uncover YSO populations. 
For the detection of diskless YSOs, i.e., Class~III sources, X-ray observations complement the 
IR YSO sample to have a more complete census of stellar content of clusters 
\citep[e.g.,][]{pre05,pan13,pand14}. The Massive Star-Forming Regions Omnibus X-ray Catalog \citep[MOXC,][]{tow14} detected X-ray sources in a sample  of massive star-forming regions, including IC~1805. 
They characterized the X-ray properties of 647 stars in the IC~1805 using $Chandra$/ACIS observations.
ACIS-I consist of four 1024 $\times$ 1024 pixel$^2$ CCD chips that cover a 17$^{\prime}$ $\times$ 17$^{\prime}$ field of view, and 
covers the 0.5 - 8.0 keV energy band with a spectral resolution of $\sim$ 150 eV at 6 keV and a PSF radius of 0\arcsec.5 within $\sim$ 2\arcmin~ of the on-axis position, 
degrading to $\sim$ 6\arcsec~ at a 10\arcmin~ off-axis angle.
We checked these $Chandra$ X-ray sources for the counterparts of our optical dataset with a less stringent  
matching radius of
3 arcsec \citep{pan13} because of the large off-axis  beam. The agreement between X-ray and optical positions 
is excellent in majority of the cases ($\sim$ 80\%), with offsets $<$1\arcsec. In a few cases, where there was more than one source within the matching radius, we considered the 
closest one as the best match. This search revealed possible optical counterparts to 403 (62 \%) of the X-ray sources. 
\begin{figure}
\centering
%\vspace{-1.1cm}
\includegraphics[scale = 0.44, trim = 10 20 10 20, clip]{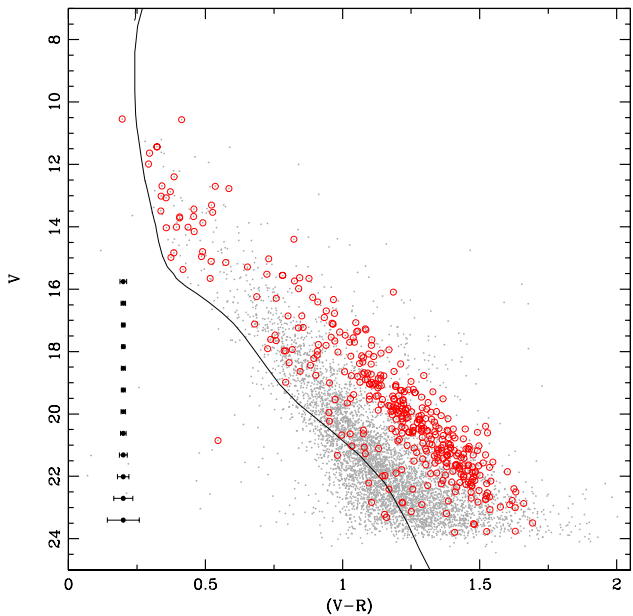}
\caption{  $V$/$(V - R)$ colour-magnitude diagram for the stars in the region covered by X-ray observations (grey dots). The MS from 
\citet{gir02}, corrected for the adopted distance and reddening are also overplotted. 
The average error in $(V − R)$ colour as a function of magnitude is shown in the left-hand side. 
Red circles represent the X-ray sources in the region.}
\label{cmd1}
\end{figure}

Fig. \ref{cmd1} shows $V$ vs. ($V - R$) colour-magnitude diagram (CMD) for the stars in the area covered by X-ray observations.
The MS from \citet{gir02} for the solar metallicity is also plotted. \citet{joshi83} have found that in the direction of IC 1805, 
$E$($B - V$) $=$ 0.7 -- 0.9 mag, indicating a small variation of extinction 
within the cluster in agreement with the A$_V$ $=$ 2.2 -- 2.7 mag, 
estimated by \citet{str13} from the 39 members of the cluster. The MS is shifted for the adopted distance of 2.0 kpc and the 
minimum reddening $E(B - V)$ of 0.7 mag. Our approach of using the minimum reddening is consistent with the  
studies of open clusters \citep[e.g.,][]{sun04,shar06,pan13}. The optical counterparts of the X-ray sources are shown with red circles. 
In Fig.3, at the low-luminosity end, one can notice a few X-ray sources scattered close to MS. Similar distribution of X-ray sources close to MS has also been observed in the CMD of 
other young clusters. For example, the distribution of X-ray sources on the $V$ vs. ($V - I$) diagram of
NGC~6530 \citep[median age $\sim$ 2.3 Myr and mean E(B-V)=0.35][]{pri05}, shows 90\% of 
PMS stars are within 0.3 -- 10 Myr, and $\sim$ 10\% of the X-ray sources are  outside 
this age range. \citet{pri05} argued that the latter sources are mainly contaminants. 

\citet{mar00} have discussed that the coronal activity 
of X-ray active field stars such as foreground dwarf M star (dMe) and background giants could lead to misidentification as young stars.
For example, in Fig. \ref{cmd1}, a few X-ray emitting sources 
are falling to the left or close to MS. Based on spectroscopic observations of the MS stars, \citet{mas95} have 
inferred that the cluster is likely to be younger than 5 Myr. Considering the young age of the cluster, 
these sources are unlikely
to be part of the cluster.
 Similarly, extended extragalactic objects seen along the line of sight could also be misidentified as 
young stars \citep{get06}. Since we have performed PSF photometry on our optical images, we expect that
the contamination due to extended objects is minimal in our sample; however we can not 
rule out the possible presence of a few point-like extra-galactic 
objects such as quasars. The field star contamination in our sample
is quantified in Sect. 3.1.5, but at least quantitatively, it appears to be 
consistent with the amount of foreground X-ray contamination expected in the direction of IC 1805.

To remove the X-ray active field stars from our X-ray YSO sample, we 
considered only those X-ray sources which are located in the empirical YSO zone 
on the $V$ vs. ($V - R$) CMD \citep[e.g.,][]{all12}. To determine the YSO zone, we used the following process iteratively. 
We arranged the X-ray sources 
in 0.75 mag bins between $V$=15.0 to $V$=24 mag. The median and standard deviation ($\sigma$) of the 
($V - R$) colour of the YSOs in each bin were calculated and the YSOs having colours greater 
than 3$\sigma$ from the median were considered as outliers. The median colour and $\sigma$ were recalculated. 
The lower and upper limits of the YSO zone in each bin
 are then taken to be 1.25 $\sigma$ and 2 $\sigma$ around the median colour of the bin. These upper
 and lower limits are then fit with polynomials and the YSOs lying within these 
upper and lower bounds are taken to be probable members. 
As discussed in \citet{all12}, the lower limits of the magnitude bins are taken closer
to the median value because the density of sources quickly
increases in the direction perpendicular to the isochrone towards the MS. In particular, foreground dwarfs
and background giants may disguise as YSOs by appearing
more luminous for a given colour. We note that although the above statistical approach is 
likely to remove non-YSO X-ray sources, some actual YSOs may also be eliminated. 
For example, a few X-ray sources located near the MS  
(where majority of the field stars are populated), 
could be YSOs, although their chances of being so are low.
Spectroscopic observations are needed to ascertain the nature of such sources,
because some of them could be YSOs with disk systems seen edge-on that appear under luminous for 
their colours because of scattering \citep{sici05} or members 
that are significantly older than the mean cluster age. 

Using the above approach, we obtained 337 X-ray emitting YSO candidates in the region covered by $Chandra$ 
X-ray observations (for details; see the Section 3.2.2). Of these, 
40 sources have shown the characteristics of Class~I/II candidate YSOs on the basis of  
2MASS and {\it Spitzer} colours (discussed in Sections 3.1.1, 3.1.2, 3.1.3), so the rest  297 X-ray 
sources are considered as Class~III candidates.   

\subsubsection{Data completeness and contamination}

\begin{figure}
\centering
\includegraphics[scale = 0.49, trim = 0 0 10 0, clip]{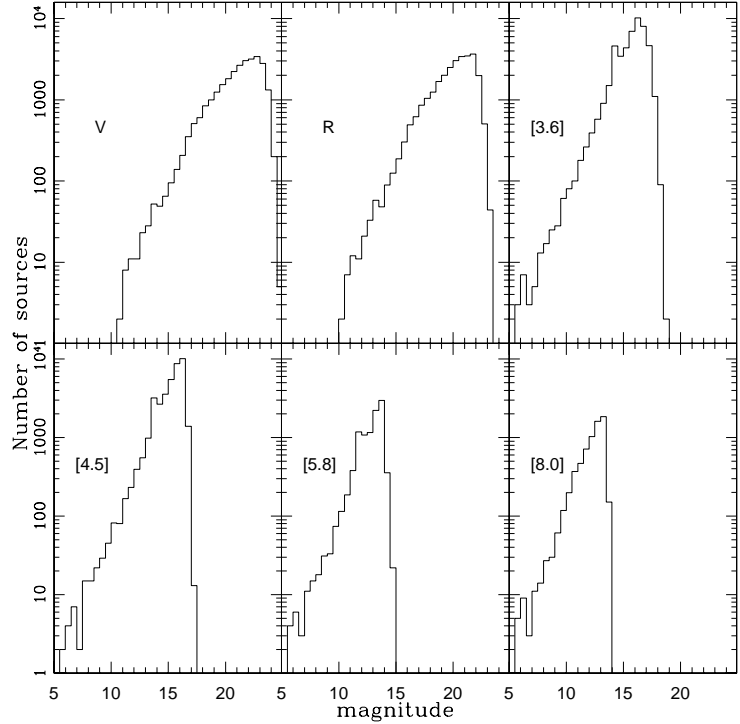}
\caption{ Histograms for the sources detected in optical (V,R) and IRAC bands showing the limiting magnitude and completeness limit for each band.
}
\label{lf}
\end{figure}

 As in the c2d legacy project \citep{eva03}, the peak of the observed luminosity function can be 
 assumed to estimate $\ge$ 90\% completeness limit \citep[see also][]{jose16}. We constructed the 
 luminosity functions for each band by using the histogram method (see Fig. \ref{lf}). 
 The resultant completeness limits of the optical and 
IRAC data are given in 
Table \ref{tab3}. 
\begin{table}
\centering
\caption{Completeness limit of the data in IRAC and optical bands.}
%\hspace{-2.0cm} 
\begin{tabular}{ccccccc}
\hline
Wavebands& $V$ & $R$ & [3.6] &[4.5]& [5.8]& [8.0]  \\
\hline
Completeness (in mag) &22.5 & 22  & 16    & 16   & 14 &13.5 \\
\hline
\end{tabular}
\label{tab3}
\end{table}

 For cross-checking, the completeness limits of $V$ and $R$ bands were also estimated by the 
artificial star experiment by using ADDSTAR routine in $IRAF$ \citep[e.g.,][]{pan13a,jos13}.
We randomly added artificial stars of various magnitudes into each image. The luminosity distribution of 
artificial stars was chosen 
in such a way that more stars were inserted towards the fainter magnitude bins. 
The frames were then re-reduced by using the same procedure as used for the original frames (see Section 2.1). 
The ratio of the number of stars recovered to those added in each magnitude interval gives the completeness 
factor  as a function of magnitude.
We thus obtained $>$90\% completeness of the photometry to be 22.5 and 22 mag in $V$ and $R$ bands, 
respectively, which agrees with that of the 
histogram analysis.

The estimation of the completeness census  of the YSO sample obtained by various colour combinations is rather difficult as it is 
limited by several factors. For example, bright extended nebulosity in the IRAC bands 
significantly limits the point source detection. The YSO identification from the 2MASS/IRAC colours is limited 
by the sensitivity of the 2MASS survey, while that from the IRAC/MIPS colours suffers from the significant saturation in the 
IRAC 8 $\micron$ and MIPS 24 $\micron$ images caused by the central luminous sources as well as the bright nebulosity. Similarly, 
the variable reddening and stellar crowding characteristics across the region could also affect the local completeness limit.
All these effects  are difficult to quantify. Since 3.6 and 4.5 $\micron$ bands are the most commonly used colour combination for identifying Class II YSOs, 
we calculate the approximate completeness limit of our Class II YSOs by using the  completeness limit of 
these two bands. The 90\% photometry completeness limits of 3.6 and 4.5 $\micron$ 
bands are estimated to be $\sim$ 16 mag. Assuming a distance of 2.0 kpc for IC 1805 and 
an average extinction A$_V$ of $\sim$ 2.5 mag, we find the photometric completeness limit 
corresponds to an approximate stellar mass in the range  0.2-0.3 M$_\odot$ for a YSO 
of age $\sim$ 2-3 Myr \citep[using the evolutionary tracks of][]{sie00}.
The identification of Class III sources in our sample is primarily based on their 
detection in both the X-ray and optical bands.
Only 62\% of the X-ray sources have optical counterparts. This indicates that 
the completeness limit of Class III YSOs is primarily 
dictated by the sensitivity limit of the optical data. The 90\% photometry completeness limits of 
V and R bands lead to  
the completeness limit of our Class III YSOs as $\sim$ 0.3 M$_\odot$ at the assumed distance,
mean $A_V$ and age $\sim$ 2.5 Myr \citep[using the evolutionary tracks of][]{sie00}. 
Adopting different sets of evolutionary tracks would provide different values of stellar masses. 
However, for low-mass objects, the tracks of \citet{sie00} are close to those of
\citet{bara98}. The agreement between masses of these two models is within 10-20\%. 
In summary, we considered that our YSO sample is expected to be largely 
complete above 0.3 M$_\odot$.

In the case of IC 1795, based on \citet{fra06} measurements from the IRAC/GOODS sample, 
\citet{roc11} calculated that the contamination of extragalactic sources in IRAC data is about 72 sources 
down to 15.5 mag at 3.6 $\micron$  for an area 0.26$^{\circ}$ $\times$ 0.26$^{\circ}$. IC~1805 is at 
the same distance and is associated with the same cloud complex as IC 1795, so 
similar fraction of extragalactic contamination is expected in  our IRAC catalog. 
Scaling their values to our cluster area (i.e., the region covered by most of our observations; see Sect. 3.1.6), we expect $\sim$ 20 extragalactic IRAC sources. 
In addition to these, other contaminants such as broad-line AGNs, unresolved knots of shock emission 
and faint sources contaminated by  copious PAH nebulosity (expected to be  
prevalent in distant massive star-forming regions) are also likely to affect our YSO selection. 
Using  the criteria of \citet{gut09}, we have removed these contaminants from our 
catalog. Although \citet{gut09} criteria for a region at $\sim$ 2 kpc may provide an overestimation of 
the contamination, but this would ensure the high reliability of our YSO sample.
Our X-ray YSO sample is selected based on the 
X-ray and optical observations. The majority of the extragalactic contaminants 
and distant background stars are expected not to have optical (V-band) counterparts.  
Moreover optical images being at high spatial resolution and with our PSF photometry, 
extended objects like galaxies are unlikely to be the major contaminants in our sample. 
In the optical CMD, most of the likely members are located in the PMS zone and 
 there are  $\sim$ 40 X-ray sources located near the MS. As discussed in Sect. 3.1.4, 
these sources are likely foreground sources of the region, and seem to agree well with 
the expected foreground contamination in the direction of IC 1805. So overall,
we expect that the likely number of X-ray contaminants in our YSO sample should 
be low.

 \subsubsection{Final catalog of YSOs}
Our final catalog includes the YSO candidates identified from $Spitzer$ IRAC/MIPS, 2MASS and X-ray datasets. 
 In total, we have 297 Class~III, 8 transition disk and 272 Class I/II candidates  
in the direction of W4 complex. However, there are only 101 Class~I/II and 283 
 Class~III candidates in the cluster (i.e., within $\sim$ 9$\arcmin$; see Section 3.2.1). 
The cluster radius is well within the common area covered by the X-ray, IRAC-MIPS as well as optical observations and our main aim is to
 explore and characterize the young stellar population associated with the cluster IC 1805.
 To determine the cluster properties, such as age (see Sect 3.2.2), mass function, disk-fraction and disk evolution of YSOs 
(discussed in sections 4.1,4.2, and 4.3), we have used YSOs within the cluster radius, 
whereas to get an idea on the large scale star-formation history, we have used all the YSOs identified in this work (discussed in Sect. 4.4). 
A sample list of the YSO candidates with their magnitudes
in different bands for the cluster area is given in Table \ref{tab2}. The entire table  is available 
in electronic form only.

In the W4 complex, using Bayes classifier and combining several criteria
in a probabilistic approach, \citet{bro13} defined the list of MYStIX Probable Complex Members (MPCM). 
Within 9 arcmin of the  cluster center, \citet{bro13} identified 389
YSOs (Xcl flag=2).  
Although  a direct comparison with 
\citet{bro13} is not possible because of different methods used to extract 
photometric magnitudes on the $Spitzer$ images (aperture photometry by them versus 
PRF photometry by us), and in the approach of YSO classification and contamination removal 
\citep[for details see][]{bro13}. Nonetheless we find reasonable agreement
between both the works. For example, within the cluster area we obtained 384 sources, 
whereas \citet{bro13} have identified 389 sources and only $\sim$ 15\% sources are 
missing in either of the catalog.

In this work, we are more interested in studying the optical properties of the YSOs 
associated with the cluster. Unlike, NIR colours, the optical filters are 
more sensitive to the temperatures of late-type stars. Thus, in the 
low-extincted regions like IC 1805, the optical CMDs are expected to be more 
reliable in separating PMS members from the field stars and also better
for the characterization of YSOs.
\begin{table*}
\centering
\caption{YSOs from 2MASS, IRAC/MIPS and X-ray data for the cluster area. The entire table is available in an electronic form. }
\tiny
\hspace{-2.0cm} 
\begin{tabular}{ccccccccccc}
\hline
Id     & RA & DEC & J$\pm$eJ & H$\pm$eH & $K_s$$\pm$e$K_s$& [3.6]$\pm$ & [4.5]$\pm$ &[5.8]$\pm$ &[8.0]$\pm$&[24]$\pm$ \\
       & (J2000) &(J2000)&   &          &         &  e[3.6]    &  e[4.5]    & e[5.8]    &   e[8.0] & e[24]   \\
\hline
     1 & 38.240929& 61.374401& 13.63 $\pm$ 0.04& 12.65 $\pm$ 0.02& 11.99 $\pm$ 0.02& 12.38 $\pm$ 0.01& 10.34 $\pm$ 0.01 & 9.58 $\pm$ 0.01&  8.57 $\pm$ 0.01&  6.94 $\pm$ 0.02\\
     2 & 38.182259& 61.469330& 12.15 $\pm$ 0.02& 11.65 $\pm$ 0.04& 11.02 $\pm$ 0.03&  9.58 $\pm$ 0.01&  8.88 $\pm$ 0.01 & 8.03 $\pm$ 0.01 & 6.72 $\pm$ 0.01&  4.15 $\pm$ 0.01\\
     3 & 38.464951& 61.406620& 14.31 $\pm$ 0.04& 13.88 $\pm$ 0.05& 13.65 $\pm$ 0.04& 13.61 $\pm$ 0.01& 13.39 $\pm$ 0.01 &13.02 $\pm$ 0.02 &12.48 $\pm$ 0.03&    - \\
.&.....&.....&.....&.....&.....&.....&.....&.....&.....&....\\
\hline
\end{tabular}
\label{tab2}
\end{table*}

\subsection{Properties of the Cluster IC 1805}

\subsubsection{Physical Extent of the Cluster}\label{rden}
\begin{figure}
\centering
\includegraphics[scale = 0.39, trim = 0 0 10 0, clip]{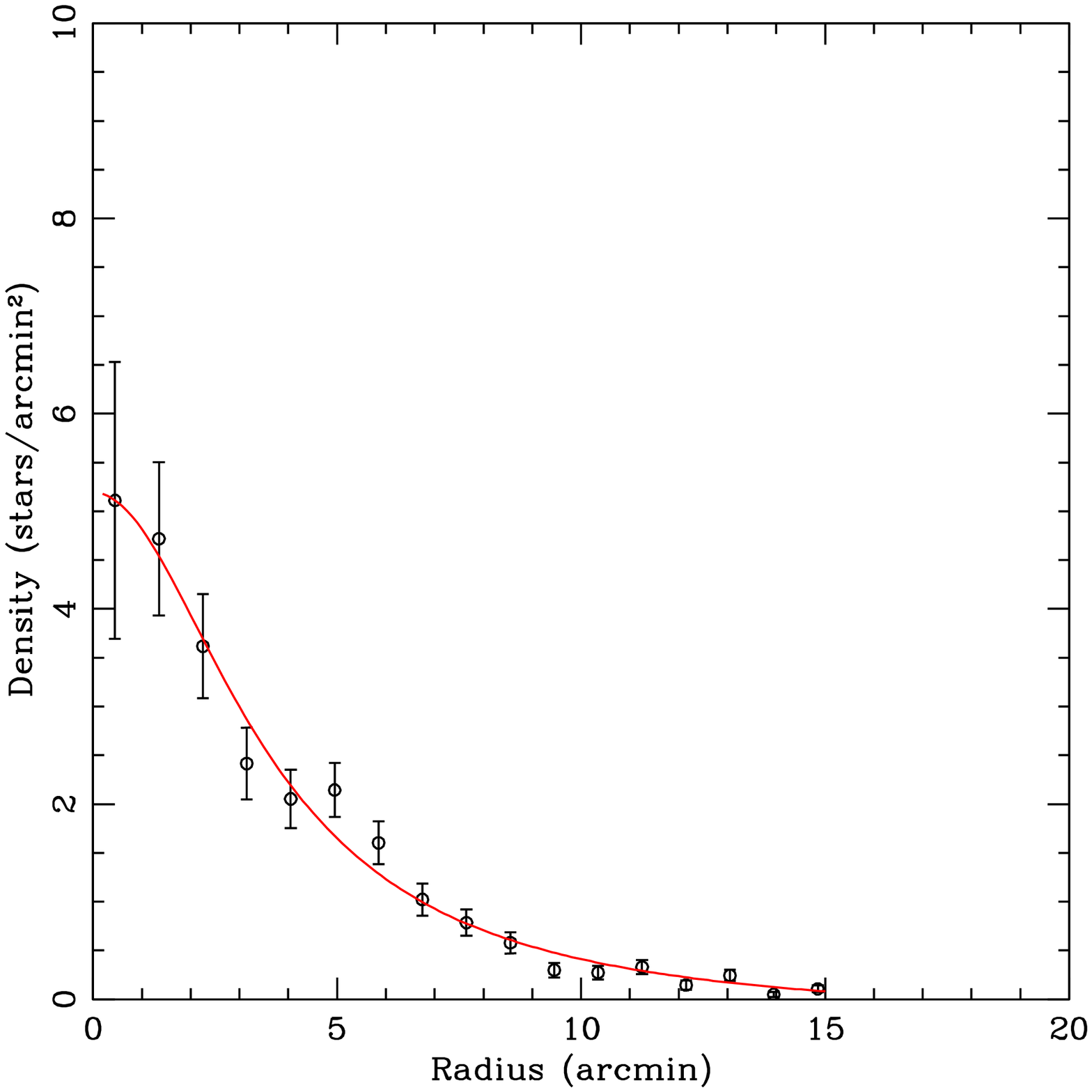}
\includegraphics[scale = 0.39, trim = 0 0 10 0, clip]{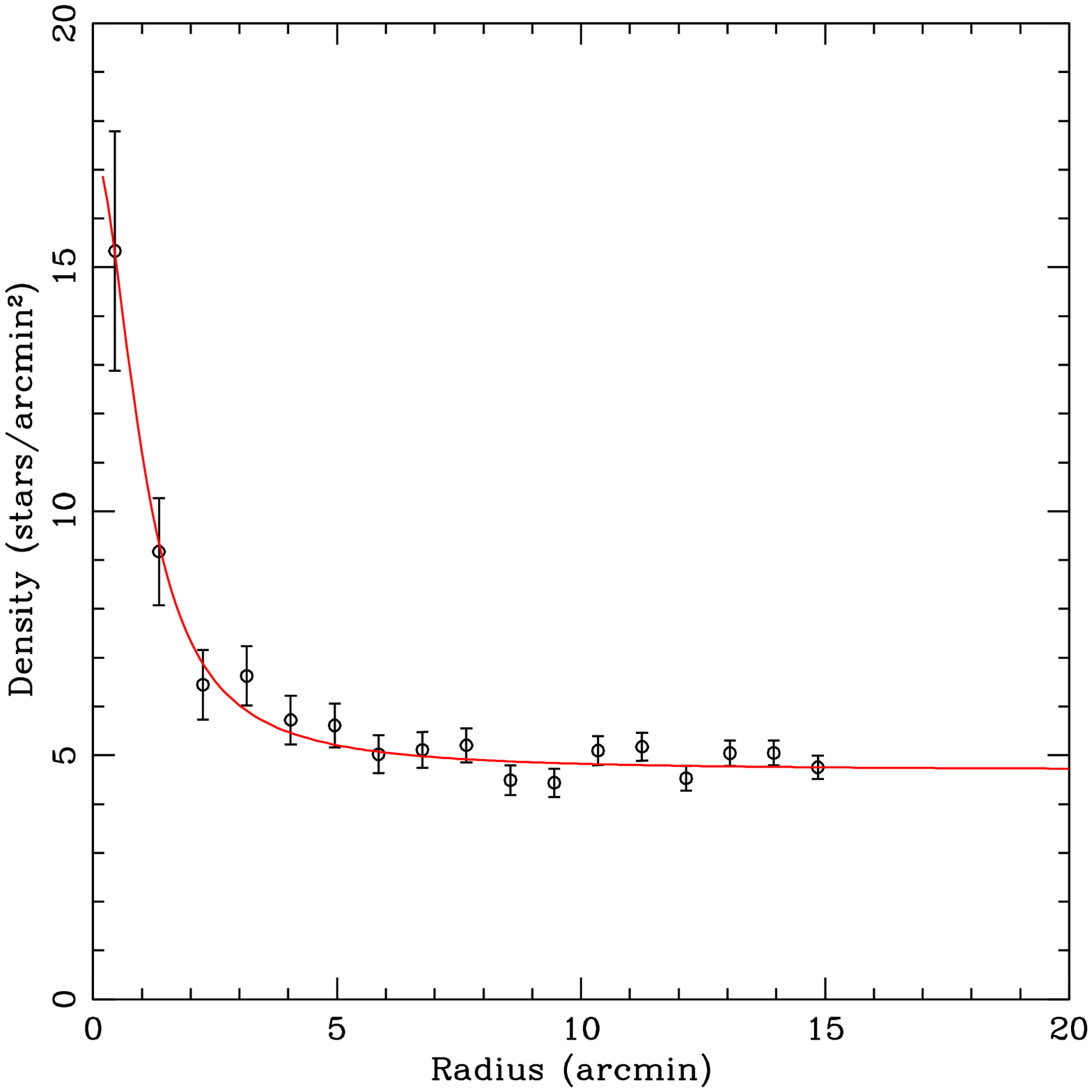}
\caption{Radial density profile of the cluster IC~1805 based on the YSO sample (left panel) and 2MASS data (right panel). 
The continuous curve shows the least-squares fit of the \citet{kin62} profile to the observed data points. The error bars represent $\pm$$\sqrt{N}$ errors. 
}
\label{rdp}
\end{figure}
The radial extent is one of the important parameters to study the dynamical  
properties of clusters. To estimate the cluster radius of IC~1805, we applied the star count technique 
and assumed a spherically symmetric distribution of stars in 
the cluster.

The point of maximum stellar density ($\alpha_{2000}$ = $02^{h}32^{m}42^{s}$; 
$\delta_{2000}$ =
$+61^{\circ}27^{\prime}21^{\prime\prime}$) taken from the $WEBDA$\footnote{https://www.univie.ac.at/webda/} is
considered as the centre of IC~1805. 
It is worthwhile to mention that the adopted cluster centre is very close to the location of most massive star of the region (see Sec. 4.3). 
We then created the radial density profile (RDP) using our YSOs catalog to study the radial structure of the cluster. 
We divided the region into a number of concentric 
circles. The projected stellar density in each concentric annulus was obtained by dividing the 
number of stars by the respective annular area. The densities thus obtained are plotted as a function of radius in Fig. \ref{rdp}. 
The  error bars are derived by assuming Poisson 
statistics. From the RDP, the cluster radius appears to be $\sim$ 9$^\prime$ (5 pc).

For comparison, we also estimated the extent of the cluster ($r_{cl}$) using 2MASS data. Within uncertainties, the $r_{cl}$ obtained from the YSO sample and 2MASS data are matching. 
The RDP is parametrised with the 
empirical model of \citet{kin62}, 
\begin{center}

$\rho (r) \propto {\displaystyle{\rho_0} \over \displaystyle{1+\left({r\over r_c}\right)^2}}$,  

\end{center}
where $r_c$ is the core radius at which the surface density $\rho(r)$ becomes half of the central 
density, $\rho_0$. The best fit (solid line) to the radial density obtained by a $\chi^2$  minimization 
technique is shown in Fig. \ref{rdp}.  
  
\subsubsection {Colour-Magnitude Diagram of YSOs : Age and Mass Estimation}
\begin{figure}
\centering
%\vspace{-1.1cm}
%\includegraphics[scale = 0.44, trim = 10 20 10 20, clip]{cmd_irac_clu2.eps}
\includegraphics[scale = 0.44, trim = 10 20 10 20, clip]{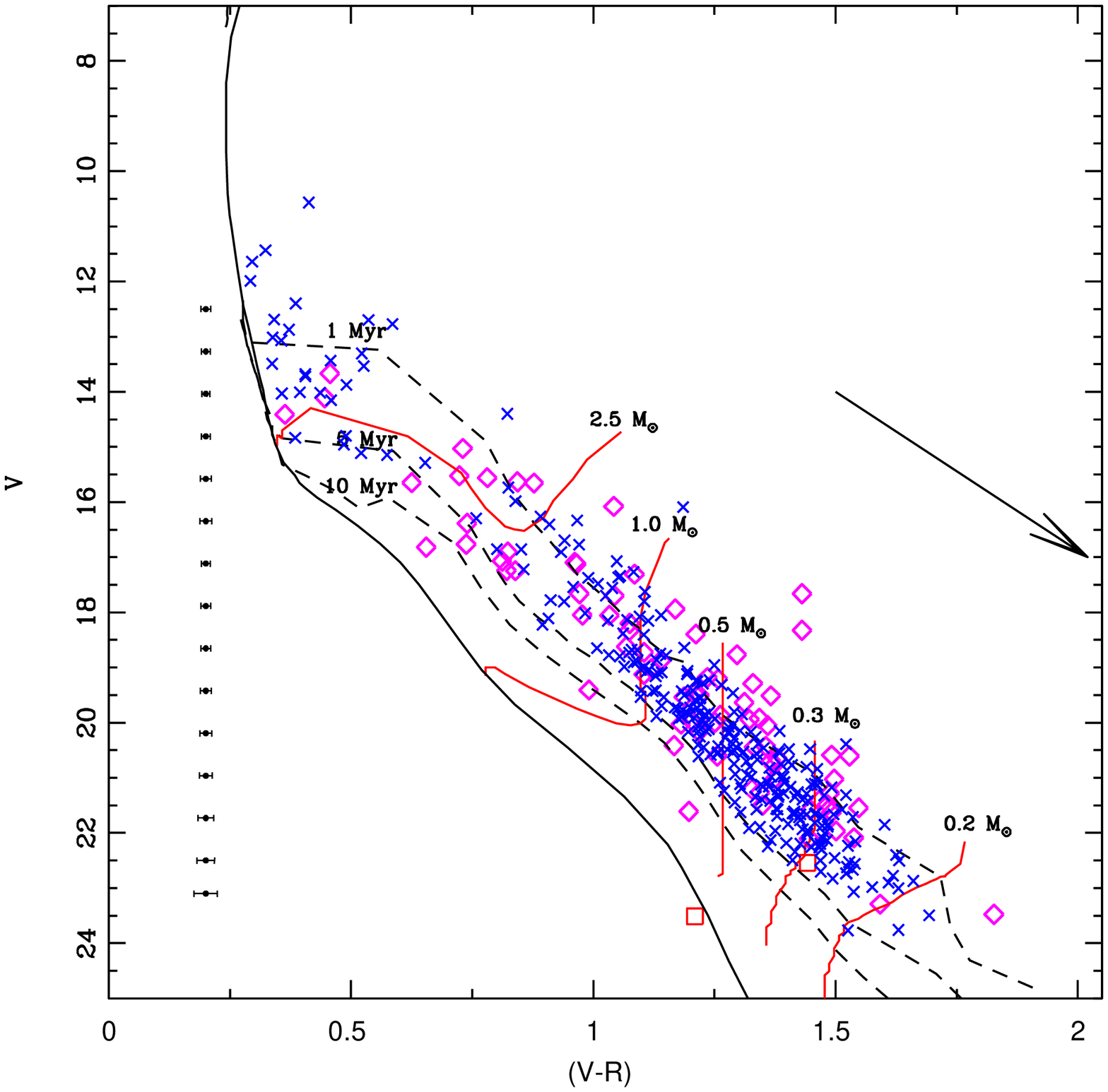}
\caption{  $V$/$(V - R)$ colour-magnitude diagram for the YSO candidates in the cluster region. The PMS isochrones from \citet{sie00} and MS from 
\citet{gir02}, corrected for the adopted distance and reddening are also overplotted. The arrow is a 
reddening vector corresponding to A$_V$ = 3 mag. Diamonds and squares represent the Class~II and Class~I sources, respectively.  
Blue crosses represent the X-ray sources which are considered as YSOs (see Sec. 3.1.4 and Sec. 3.2.2) in the present study.}
\label{cmd}
\end{figure}
 The ages and masses of the cluster
members were estimated by comparing their locations on the CMD with PMS isochrones of various ages after correcting for
the distance and extinction. The minimal differential extinction ($\Delta$ $A_V$ $\sim$ 0.6 mag) towards the cluster is an advantage to examine the position of the YSOs on the CMD in order to confirm their youth,
to determine their optical properties and to study the star formation history of the region. 

Fig. \ref{cmd} shows $V$ vs. ($V - R$) CMD for the YSOs in the cluster area.
The MS from \citet{gir02} as well as the PMS isochrones for 1, 5 and 10 Myr for the solar metallicity from \citet{sie00}, corrected for the 
adopted distance and reddening are also overplotted.    
The Class~I and Class II sources (squares and diamonds, respectively) within the cluster area are identified based on the $Spitzer$ IRAC/MIPS photometry. 
The location of X-ray emitting (likely Class III) sources are shown with the crosses. 
The age distribution of the majority of these sources is in 
the range of 0.5 - 7 Myr, concentrated around an age in the range of 2-3 Myr
 (mean age 2.5 $\pm$ 1.5 Myr). We find that masses for a majority of the YSOs are in the range of  0.2 M$_\odot$ -- 2.5 M$_\odot$. 
As discussed in Sect. 3.1.4, our approach to select the Class III candidates might have removed some YSOs, which may affect the mean age of 
the cluster. Hence, we also estimated the age of the cluster using only the disk-bearing sources 
(i.e, sources that have shown excess emission in the 2MASS and IRAC bands). This method 
resulted an age $\sim$ 2.3 Myr for the cluster, which is comparable to the mean age estimated using
all the YSOs. 

Age spread is common in young clusters and is probably 
due to the combined effect of differential reddening, circumstellar extinction, variability, 
binarity and/or different evolutionary stages \citep{jos17}. However, the reddening vector is nearly parallel to the isochrones, hence differential reddening does not have much effect 
on the age estimation of YSOs. The circumstellar extinction may affect the age and mass determination of Class 0/I sources, but its effect 
should be minimal in optically visible PMS sources such as Class III YSOs. Though the exact role of 
circumstellar disks requires detailed SED modelling or spectroscopic observations, which is
beyond the scope of this paper, it is worthwhile to mention that \citet{jos13} performed SED modeling on a set of
optically visible PMS sources and found that SED based ages are largely in agreement with 
those estimated from the optical CMDs. 
Binary companion 
will apparently brighten the star, consequently 
CMD will yield a younger age. For example, in the case of equal-mass binaries, cluster is expected to 
show a sequence in the CMD which is shifted by 0.75 mag upwards. The effect of binarity 
can be seen more prominently in some older clusters ($\sim$ 10 Myr) for which the isochrones are close together 
\citep[e.g., NGC 7160;][]{sici05} whereas in the case of young clusters like Tr 37 ($\sim$ 4 Myr), which is similar to IC~1805,
the isochrones are well separated, hence the effect of binarity on age estimation will be less compared to 
the natural age spread of the stars in clusters \citep[see discussion in][]{sici05}.
 Similarly, though TTSs tends to show variability at optical bands
 \citep[e.g.,][]{her94,bri01,rod10,lat15}, 
the variability tends to move the objects parallel to the isochrones,
resulting in little effect on the age estimation of YSOs \citep{bur05}. 
 Despite of low differential reddening and presence of large number of disk-less sources 
in the cluster, some degree of age spread due to the combination of the above factors is expected, 
but since the observed spread in colour due to member YSOs is larger than their mean uncertainties 
due to photometric colours (see Fig \ref{cmd}), we expect that the YSOs
are at different evolutionary stages.

We considered that the errors associated with the determination of age and
mass are mainly of two kinds; random errors in photometry, 
and systematic errors due to different theoretical
evolutionary tracks. We estimated the effect of random
errors by propagating them to the observed estimations of $V$, ($V - R$) and
$E(V - R)$ by assuming a normal error distribution and using Monte Carlo simulations \citep[see e.g.,][]{cha09}.
Since we have used the evolutionary model by \citet{sie00} for all
the PMS stars, the age and mass estimations given in Table 4 should not
be affected by systematic errors. A sample of Table~4 is given here 
and the complete table is available in the electronic version. 

\begin{table*}
\centering
\caption{Magnitudes, age and mass of the YSOs in the cluster region. Complete table is available in an electronic form. }
\begin{tabular}{ccccccc}
\hline
\hline
Id     &RA (J2000) & DEC (J2000)&             V $\pm$ eV & R $\pm$ eR        & Age $\pm$ error Age (Myr)     & Mass $\pm$ error in Mass (M$_\odot$) \\
\hline
\hline
     1&  37.888229& 61.429279& 22.10 $\pm$ 0.02& 20.66 $\pm$ 0.01&  3.01 $\pm$ 0.65&    0.31 $\pm$ 0.02\\
     2&  37.895050& 61.506271& 21.28 $\pm$ 0.01& 19.79 $\pm$ 0.01&  1.20 $\pm$ 0.30&    0.30 $\pm$ 0.02\\
     3&  37.919338& 61.502380& 19.87 $\pm$ 0.01& 18.61 $\pm$ 0.01&  1.57 $\pm$ 0.57&    0.57 $\pm$ 0.08\\
... &...&...&...&...\\
\hline
\end{tabular}
\label{tab4}
\end{table*} 

\section{Discussion}
 \subsection{Mass function of the YSOs }
\begin{figure}
\centering
\includegraphics[scale = 0.40, trim = 0 0 0 0, clip]{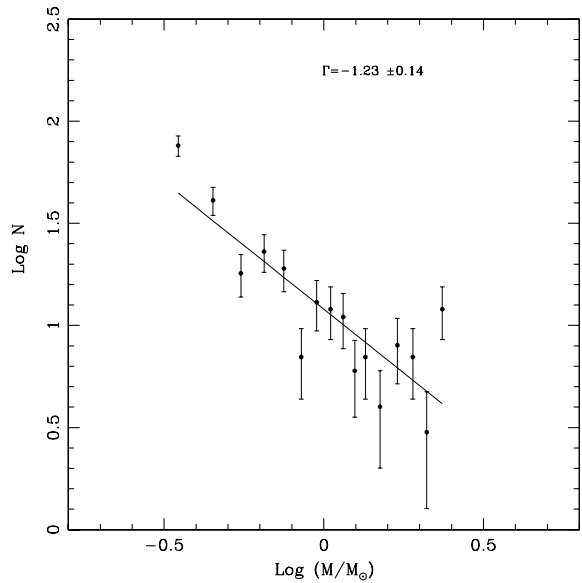}
\caption{The mass function of the candidate YSOs in the mass range (0.3$\le$ M/M$_\odot$ $\le$ 2.5), derived from the optical data. The error bars 
represent $\pm$ $\sqrt{N}$ errors. The continuous
line shows the least-square fit to the mass ranges described in the text. The
value of the slope obtained is given in the figure. }
\label{mf}
\end{figure}

Young clusters are the best objects to study the initial MFs. Since young low-mass stars have limited time 
to segregate from the cluster and/or loose mass through evolutionary processes, the variation of the MF 
gives clues about the physical conditions of the star formation process.  
The MF is defined as the number of stars per unit logarithmic mass interval,
and is generally represented by a power law with a slope, \\
$\Gamma$ = {{ $\rm {d}$  log $N$(log $m$)}/{$\rm{d}$ log $m$}},\\
 where $N$(log $m$)
is the  distribution of the number of stars per  unit logarithmic mass interval. 
Most of the star clusters in the solar 
neighborhood have MF slopes near the Salpeter MF slope, i.e., 
$\Gamma$ = --1.35. 

Since our YSO sample is 
complete down to 0.3 M$_\odot$, for the MF study we have taken only those YSOs which 
have masses in the range 0.3 $\le$ M/M$_\odot$ $\le$ 2.5. 
The mass distribution of our YSO sample has a best-fit slope, 
$\Gamma$ = -−1.23 $\pm$ 0.14 (see Fig. \ref{mf}). 
\citet{nin95} studied the cluster IC~1805 and found a slope of $\Gamma$ = -−1.38 $\pm$ 0.19 for masses between 
2.5 and 30 M$_\odot$. A similar value of $\Gamma$ (= --1.3 $\pm$ 0.2) has been reported by \citet{mas95} for the 
massive stars of the W4 complex. 
Our YSO MF is consistent with those reported for other active star forming regions. 
\citet{eri11} derived the MF of an extinction limited YSO sample (with masses 
$>$0.2 M$_\odot$) consistent with the 
field star MF. \citet{kan09} did not find any 
difference in the YSO MF slopes of the star-forming regions W51~A ($\Gamma$ = --1.17 $\pm$ 0.26) and 
W51~B ($\Gamma$ = --1.32 $\pm$ 0.26). 
  We note that a few older Class III sources might have been excluded as contaminants due to 
 our conservative selection criteria (see Sec. 3.1.4). Even then, we believe our result should represent the IMF 
for the cluster, for example,
and low-mass isochrones are very close to each other.

\subsection{Evolution of T-Tauri Stars in IC~1805}
TTSs are low-mass stars
($<$ 3M$_\odot$) which are contracting towards the MS. They are generally classified  into weak-line TTSs (WTTSs) and classical TTSs (CTTSs) on the basis of the
strength of the H$_\alpha$ emission line \citep[][]{strom89}. WTTSs exhibit a weak, narrow H$_\alpha$
(EW $\leq$ 10 \AA) in emission with no or little infrared excess, while CTTSs display a strong H$_\alpha$ emission line 
(EW $\ge$ 10 \AA) and large infrared excesses.
Class II sources are generally considered to correspond in the optical category to CTTSs and 
Class III sources to WTTSs \citep[e.g.,][]{hart05,pan13}. 
The `standard model' by \citet{ken95} postulates that the CTTSs evolve to WTTSs by losing
their circumstellar disks or at least their inner parts. In the Taurus region  the WTTSs are systematically older
than the CTTSs \citep{ber07}, but the statistical significance is low \citep{ken95,hart01,arm03}. 
Contrary to the above results, there are studies which
favour the assumption that CTTS and WTTS are coeval and possess indistinguishable stellar
 properties \citep[][]{law96,gra05}. 
The coevality of CTTSs and WTTSs in a star-forming region can be explained by assuming 
that YSOs have intrinsically a wide range of disk masses
and their accretion activity and/or the dispersal of the disk takes
place in a correspondingly wide range of time scales \citep{fur06,ber07}. 

\begin{figure}
\centering
\includegraphics[scale = 0.35, trim = 0 0 0 0, clip]{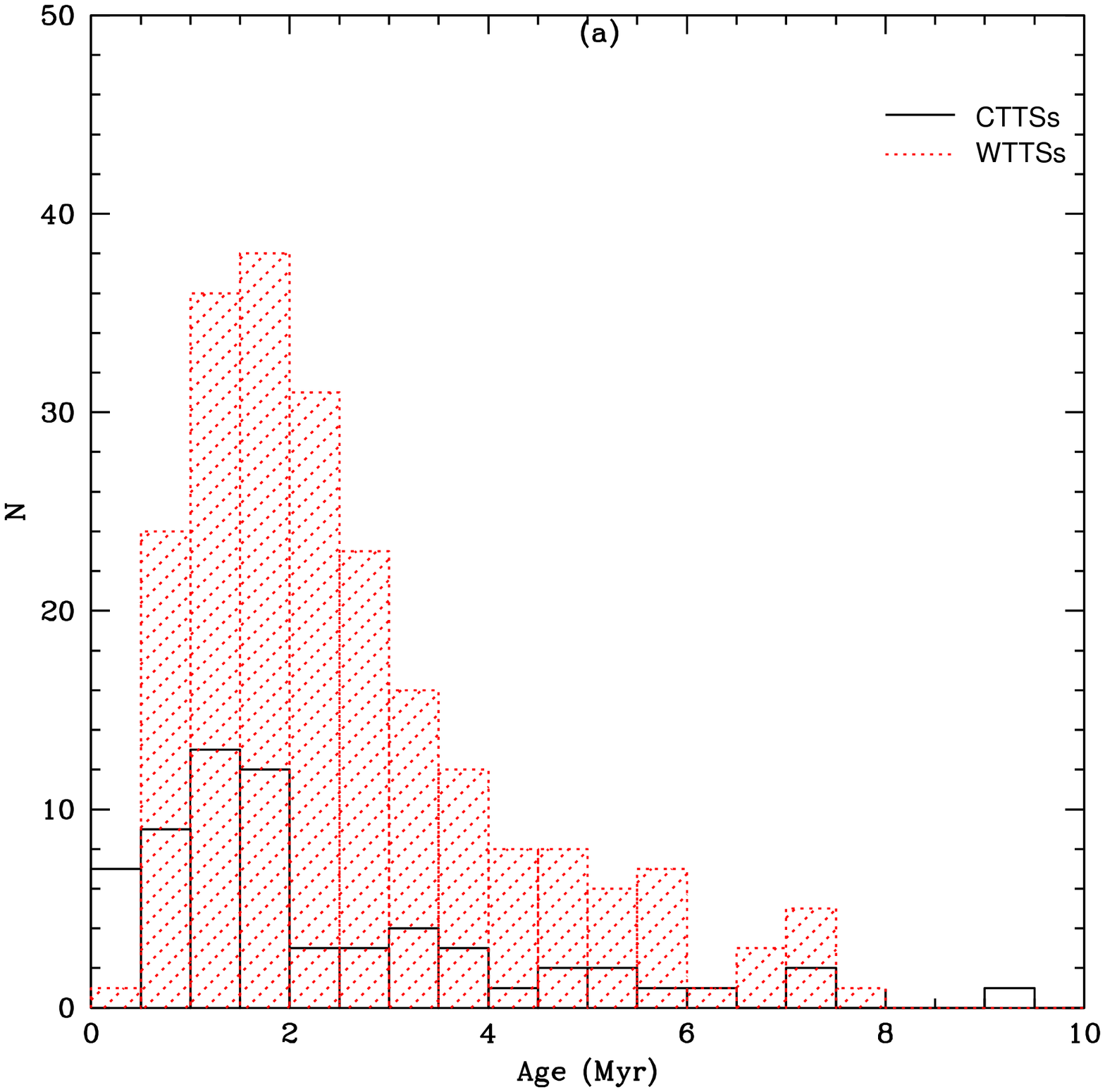}
\includegraphics[scale = 0.35, trim = 0 0 0 0, clip]{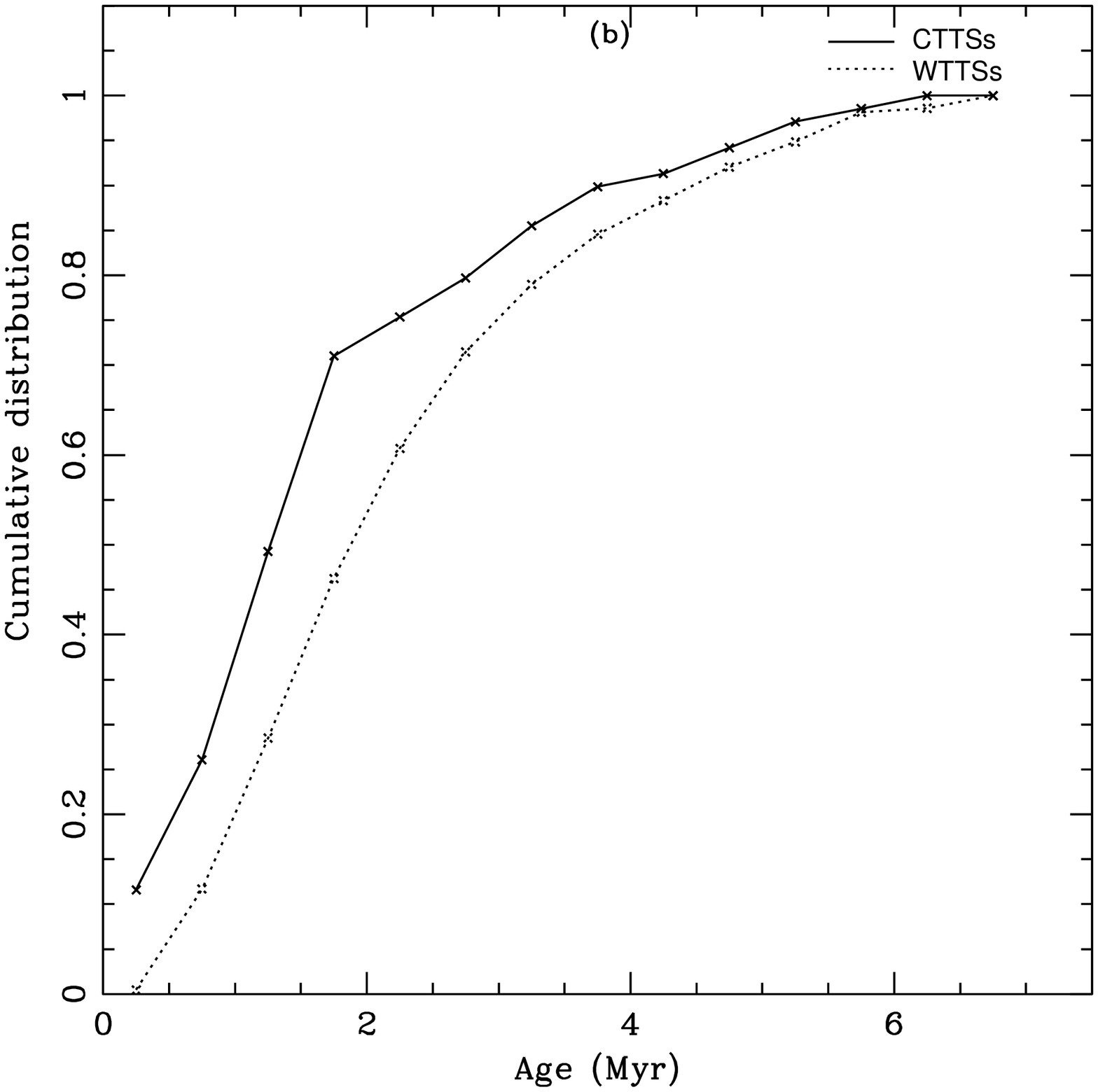}
\caption{(a) Histograms of the age distribution of the Class~II and  Class~III candidates, (b) Cumulative distributions of CTTSs and WTTSs in the cluster region as a function of stellar age.  }
\label{hist}
\end{figure}

In the present work, we have derived the ages of a large sample of YSOs in the IC~1805 cluster, 
hence, it is worthwhile to attempt to address the problem of the evolutionary status of CTTSs and WTTSs. 
Fig. \ref{hist}a shows the histograms of the ages of 
the CTTS and WTTS candidates within the cluster region in the mass range 0.3 -- 2.5 M$_\odot$. 
This manifests that the CTTSs are relatively 
younger (mean age $\sim$ 2.3 Myr) than the WTTSs (mean age $\sim$ 2.6 Myr). The cumulative distribution 
of the ages of CTTSs and WTTSs are shown in Fig. \ref{hist}b. We obtained the Kolmogorov-Smirnov (KS) test  value of 11\% that the two 
samples are drawn from the same population, which is marginally significant. 
Although the statistical significance is less, our result is in agreement with those of \citet{ber07} for the 
Taurus-–Auriga T-association and of \citet{cha09} for the BRC regions, that WTTSs are relatively older than CTTSs, 
which supports the notion that CTTSs evolve into WTTSs. Here again, we would like to remind readers 
that these results are affected by the uncertainty associated with the  membership 
of the stars (see Sec. 3.1.4), particularly the WTTSs (Spectroscopic information is needed to confirm the true nature of WTTSs and CTTSs).
\subsection{Disk Fraction Variation in the Cluster}
\begin{figure}
\centering
\includegraphics[scale = 0.44, trim = 0 0 0 0, clip]{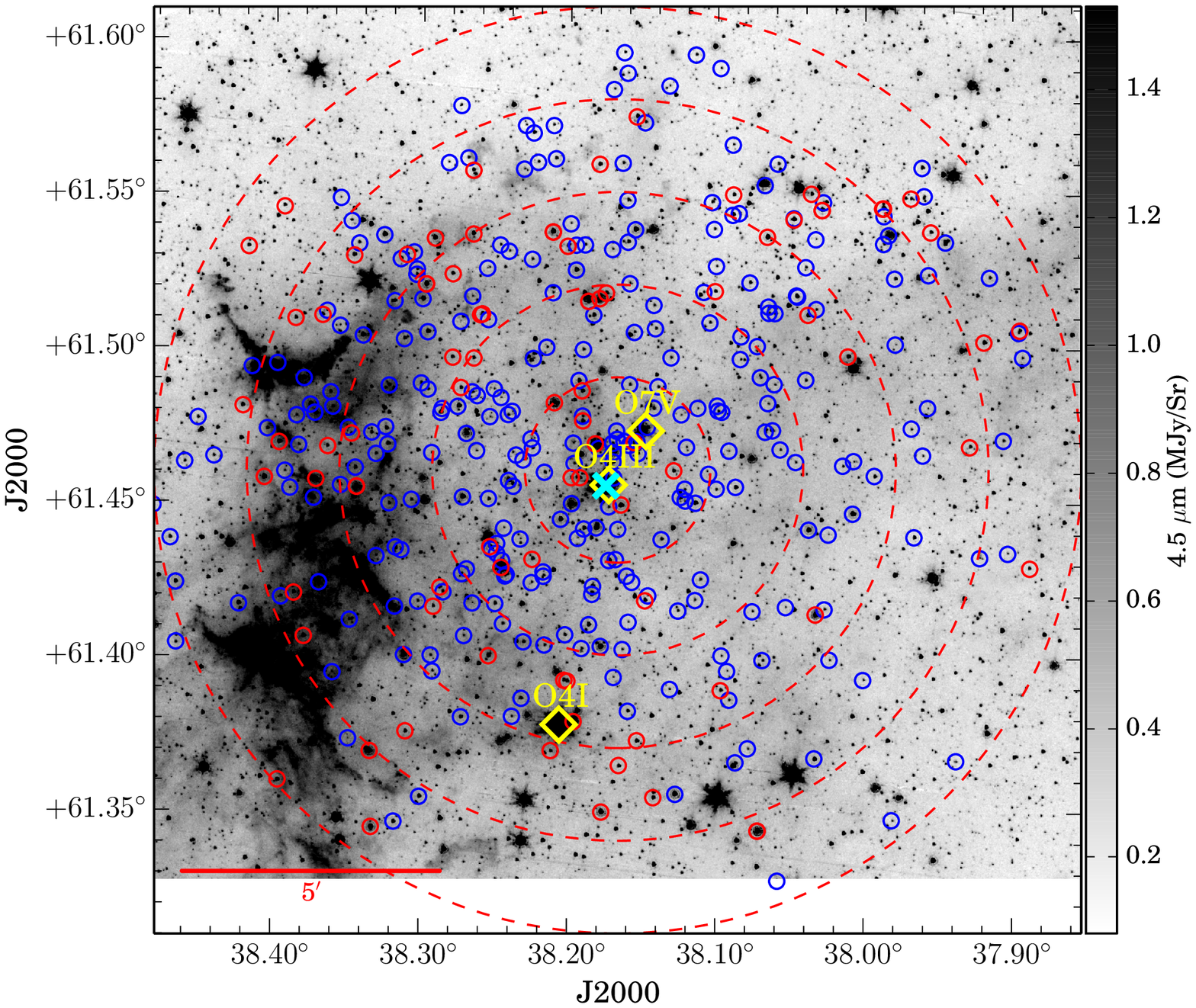}
\includegraphics[scale = 0.36, trim = 0 0 0 0, clip,angle=270]{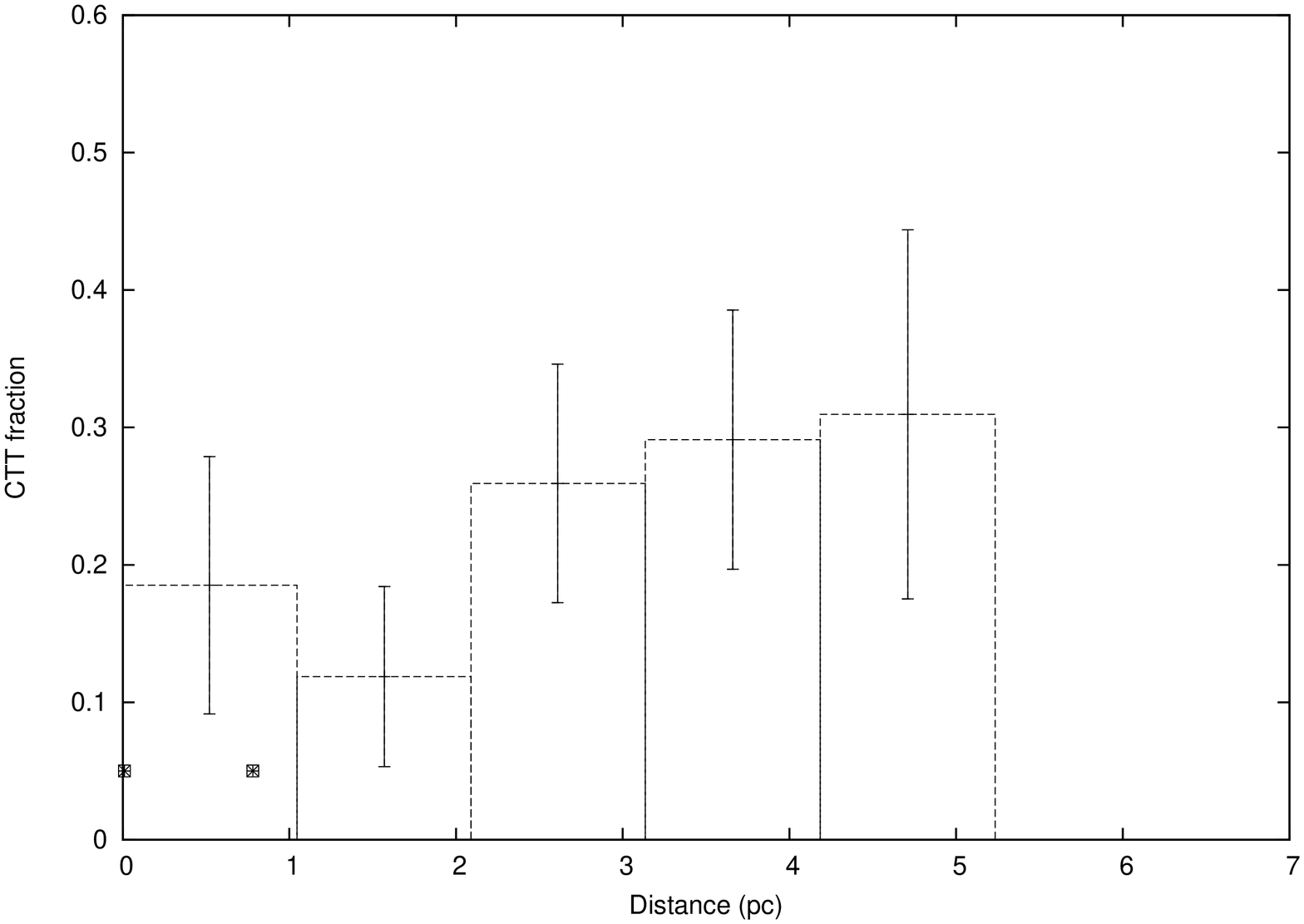}
\caption{ The upper panel shows the surface distribution of WTTSs (blue circles) and CTTSs (red circles) 
superposed on the $Spitzer$ 4.5 $\micron$ image of the cluster. The concentric annuli have a width of 1$^\prime$.8. The massive stars are shown with diamond symbols and 
cluster centre with cross. The lower panel shows the CTTS fraction as a function of the radial distance 
from the cluster centre. The two filled squares represent the locations of the two massive stars near the cluster centre.}
\label{df}
\end{figure}
The disk fraction of YSOs in clusters decreases 
with ages \citep[e.g.,][]{hai01,her08}. \citet{hai01} quantitatively
concluded that in young clusters an initially very high (80\% - 90\%) disk fraction decreases to 50\% in $\sim$ 3 Myr and 
reaches almost 10\% at $\sim$ 5 Myr. 
Generally massive stars seem
to lose their disks earlier than lower-mass stars. For example, most low-mass stars (spectral type K5 and later)
 lose their disks in 5$-$7 Myr \citep[e.g.,][]{sici05}, whereas for the intermediate-mass Herbig Ae/Be stars the corresponding 
time scale is $<$3 Myr \citep[e.g.,][]{her05}.

In a young cluster the disk fraction depends also on the
spatial position. Theoretical calculations predict that external UV radiation 
of high-mass stars can photoevaporate outer disks only within 0.3 $-$ 0.7 pc \citep[][]{john98,adam04}. 
Decrease in disk frequencies in the immediate vicinity of massive stars has been found
 in several massive clusters, e.g., NGC 2244,  NGC 6611 and Pismis 24 
\citep{bal07,mer09,fang05}, suggesting rapid destruction of circumstellar disks in such environments. 
IC 1805 is a young cluster (age $\sim$ 2.5  Myr) with several
 low-mass ($<$ 2.5 M$_\odot$) stars that have disks, 
thus the cluster represents a good site for studying the influence of massive stars on the evolution of disks. 

In  IC 1805, it is argued that the region is mainly powered by three O-type stars \citep{lef97}. BD +60502 and BD +60501 are of
spectral type O4.5III  \citep[][]{sota14} and O7V \citep[][]{sota11}, respectively and are located 
near the assumed cluster centre (see 
Fig. \ref{df}), whereas HD 15570 is an  O4I star \citep[][]{sota11}, located 
$\sim$ 4.7\arcmin~ east of the cluster centre. The radial velocities of BD +60502 and BD +60501 are --42 km/s 
\citep{khar07} and --52 km/s \citep{khar07}, respectively, consistent with 
the molecular cloud velocity ($-55$ to $-32$ km/s) associated with the W3/W4 complex 
\citep[][]{heyer88}. The radial velocity of HD 15570 is $-$24.00 km/s \citep{pul06}, 
thus unlikely to belong to the W3/W4 complex, but seems 
consistent with that of an object in an inter-arm cloud, whose  velocity lies between 
$-$32 to $-$18 km/s \citep[][]{heyer88}. We conclude that 
the cluster is dominated by the O4.5III and O7V stars 
located near the cluster center.

To study the influence of these massive stars on the evolution of nearby low-mass stars, we divided 
the cluster area into five concentric annuli (each of width 1.8$^\prime$). 
Fig. \ref{df} (upper panel) displays the distribution of CTTSs (red circles) and WTTSs (blue circles) 
in the mass range 0.3 - 2.5 M$_\odot$
from the catalogue of YSOs for which we have mass estimations from the optical CMDs.  Our aim is to examine the disk fraction variation of the low-mass 
stars within the cluster due to external influence, 
as massive stars can self-destroy their disk, because evolution of protoplanetary disks is faster 
around higher mass stars \citep[e.g.][]{ken09,yas14}. Fig. \ref{df} (lower panel) 
displays the fraction of low-mass  CTTSs 
(i.e., the ratio between CTTSs and CTTSs+WTTSs  
in each annular area) as a function 
of the projected radial distance from the cluster centre which is only 0.03 pc away from 
BD +60502. In the Figure,
the substantial uncertainties shown as solid lines originate from the low statistics of 
IR excess sources in each bin. We also checked the distribution of disk frequency by re-centering 
the analysis on the O7 star and find the disk frequency variation to be  
nearly the same. Although we see a dip in the disk-fraction vs. distance plot in the vicinity of 
the O-type star, due to large statistical error, 
we do not have a strong evidence of variation in the disk fraction within the cluster. 
However, we note that the disk fraction estimation 
can be affected by many factors, such as  various methods used to identify YSOs, 
the variation in stellar density \citep[e.g.,][]{spe15}, sensitivity of different bands and  
the uncertainty in the membership of the older WTTs. For example,
$Spitzer$ observations of high mass star forming regions
suffer from high, non-uniform nebulosity, primarily emission from PAHs at 
IRAC bands and is strong close to the massive stars near the center of 
rich clusters, where stellar crowding is high. In contrast Chandra's sensitivity 
is less affected by nebulosity and moreover it has several times better on-axis spatial 
resolution than $Spitzer$. So IRAC bands can
lead to severe decrease in sensitivity of the disk-bearing YSOs compared to disk-less YSOs
near the cluster center. Nonetheless, our result is in agreement with 
\citet{roc11}, who found no variation of the disk fraction as a
function of the distance from the high-mass stars in IC 1795, which has
 similar number of high mass stars and located in the same cloud complex.

\subsection{Star Formation History}
\begin{figure}
\centering
\includegraphics[height=13cm, width=10cm]{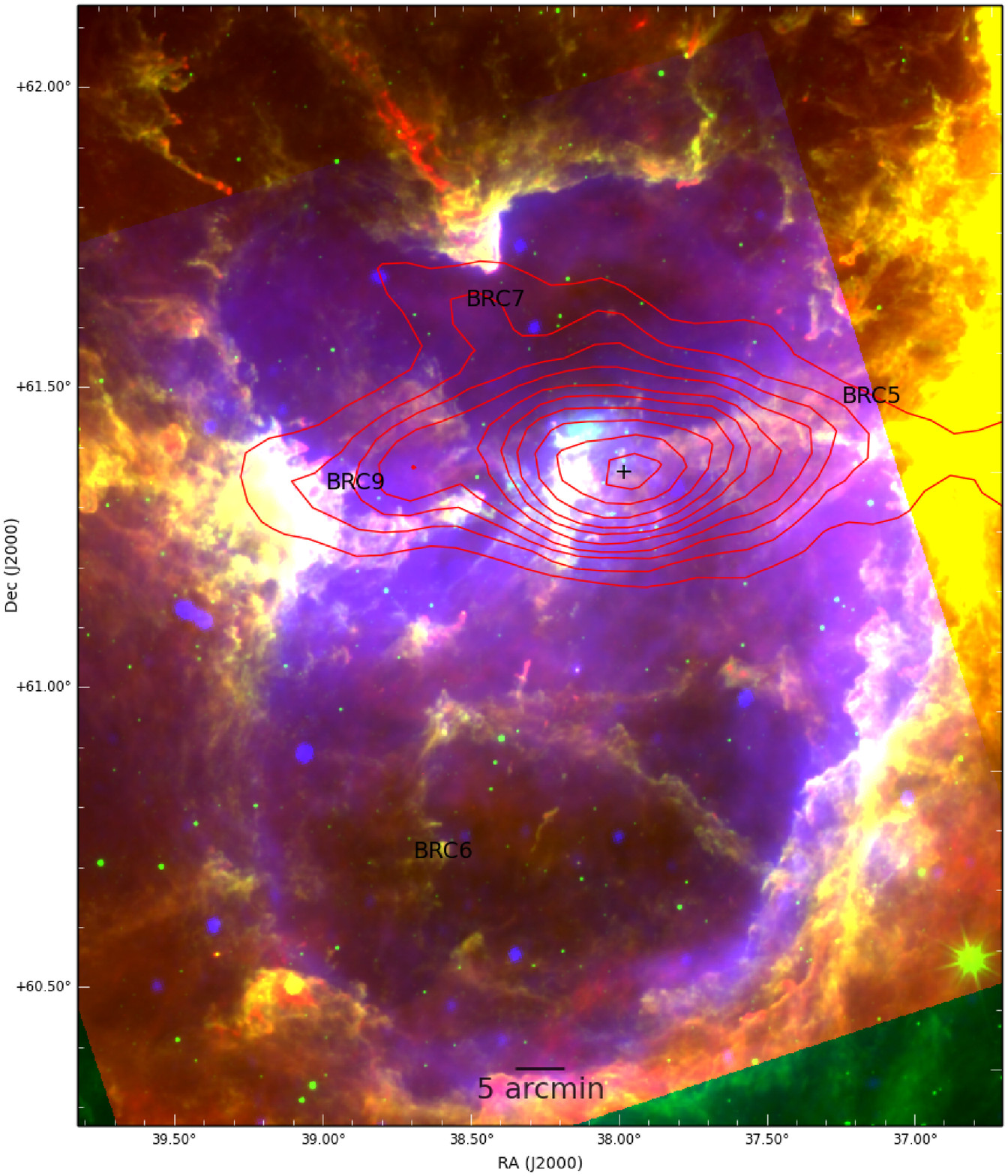}
\caption{Colour composite image of IC~1805 region reproduced with CGPS 1.4 GHz emission (blue colour) superimposed on those of the
WISE 12 $\micron$ emission (green colour) and $Herschel$ 250 $\micron$ emission (red colour). BRCs associated with the region and cataloged in \citet{sug91} are also marked in the figure. 
The red contours display the surface density distribution  
of the YSOs with the plus symbol representing the cluster centre. }
\label{figher}
\end{figure}
Figure \ref{figher} shows the Canadian Galactic Plane Survey \citep[CGPS;][]{tay03} 1.4 GHz emission
 in the direction of the W4 bubble, superimposed on the
WISE\footnote{http://irsa.ipac.caltech.edu/applications/wise/} 12 $\micron$ emission and $Herschel$\footnote{http://www.cosmos.esa.int/web/herschel/science-archive/} in 250 $\micron$ emission. We have also marked the BRCs in the region \citep{sug91}. 
The 12 $\micron$ 
WISE band  contains  11.7 $\micron$ emission commonly attributed
to PAH molecules excited in the photon dominated regions (PDRs) at the 
interface of H{\sc ii} 
regions and their adjacent molecular clouds by the far-UV photons leaking from the H{\sc ii} regions
 \citep{pom09}. Therefore, PAHs are good tracers of the warm PDR that surrounds the H{\sc ii} 
region. PAH emission is also a good tracer of newly formed, embedded B-type star formation \citep{peet04}, as these
stars heat the surrounding dust to high temperatures enough to excite the PAH bands and fine-structure lines. 
Figure \ref{figher} reveals a filamentary distribution of the 12 $\micron$ emission with its
long axis in the east-west direction that bisects the H{\sc ii} region and also in the PDRs at the 
periphery of the bubble. The  dust emission detected by the $Herschel$ 250 $\micron$
also has an elongated distribution similar to that of the 
12 $\micron$ emission. 
Compared to the PAH and  dust emission, the
ionized gas appears to be distributed orthogonal to the long axis of the
filamentary structure, with decreasing intensity away from the
filament axis. The ionized gas seems to be bounded more towards the east and west
directions than in the north and south. The overall morphology
of the nebula looks bipolar. 
Such a bipolar morphology  has been noticed
in NIR to MIR bands in a few H{\sc ii} regions and/or bubbles \citep[see e.g.,][]{sai09,sam12,mali13,deh12}. 
Recently $Herschel$ observations have clearly shown the
presence of a cold dense neutral filament in bipolar H{\sc ii} regions that bisects ionized lobes
 \citep[e.g.,][]{deh15}, similar to the one observed here.
Thus, the elongated nature of the 250 $\micron$ emission  most likely represent the long axis of 
the parental filament. 

It has been suggested that the stellar distribution in star forming regions is governed by the structure of the parental molecular 
cloud as well as how star formation proceeds in the cloud \citep[e.g.,][]{chen04,shar06,schm08, 
gut09,sam15,jose16}. Although our survey does not sample all the YSOs of the W4 complex,
it is good enough to provide clues on the star formation history.  
We generated the YSO stellar surface density map using the nearest neighborhood  method \citep{gut05}. 
At each 
sample position [i, j] in a uniform grid, we measured r$_N$(i, j), the projected radial distance to the N$^{th}$ nearest star.
The value of N is allowed to vary depending upon the desired smallest scale structures of interest. 
We generated a map using N=20, which, after a series of experiments, was obtained as a good compromise between 
the resolution and signal-to-noise ratio of the map. In the stellar surface density distribution (see Fig. \ref{figher}),
 the YSOs show a centrally concentrated clustering. At the same time, the overall distribution has an elongation along the east-west direction, 
 following the structure revealed by the 250 $\micron$ emission. Thus it appears that the formation of the cluster 
IC~1805 
possibly started in a filamentary cloud, where the density was high along the axis
of the filament and low towards the perpendicular directions.
With time the O stars of the cluster developed an H{\sc ii} region, which subsequently
grew in size, first inside the dense filament, then opening out in a hole in
the low density direction of the 
filament, resulting a large cavity in the north and south direction. Based on the distribution of
young stars on the {\it Herschel} images, \citet{deh12} have suggested that young clusters in the W5 
complex were also formed in a filamentary cloud. In the W4 complex, we could see a few BRC structures in the
north and south directions facing the IC~1805 cluster (see Fig. \ref{figher}). This could be due to
the fact that as the ionized gas streams away in the low density side,
it encounters a few small, dense clouds along its way, forming the BRC structures.

The YSO distribution along the filament axis of the W4 bubble
suggests that some of them could have resulted from the fragmentation of
the original filament as are the cases found in other filamentary systems \citep[e.g.,][]{jack10,haca11,sam15} and/or due to compression of
dense gas of the primordial filament by an expanding H{\sc ii} region. Though  precise characterization (e.g., proper motion and age) of individual
YSOs projected on the filament is needed to understand their origin, the overall morphology 
of the W4 H{\sc ii} region and the distribution of the associated 
YSOs suggest that IC~1805 could have been formed in a filamentary cloud.
\section{Conclusions}
With the aim of unraveling the less known low-mass YSO population of the cluster IC~1805 which is rich in 
massive OB stars, we studied the 
region using optical, IR and X-ray datasets. Our results suggest that despite the less favorable conditions 
for star formation around the high-mass stars, IC\,1805 shows 
signs of low-mass star formation similar to other clusters in the solar vicinity. We identified and characterized 
a large number of low-mass YSO candidates in the region. Most of them have masses in the range 0.2 - 2.5 
M$_\odot$ and age 0.1 - 5 Myr. The slope of the MF in the mass range 
0.3 $\le$ M/M$_\odot$ $\le$ 2.5 is found to be $\Gamma = -−1.23 \pm 0.14$, similar to the Salpeter MF. 
The mean age of the candidate YSOs is found to be $\sim$ 2.5 Myr. The candidate WTTSs are found 
to be relatively older when compared to the CTTS candidates. We do not find strong evidence for 
disk dispersal due to massive stars. 
 The spatial distribution of the YSOs, dust and 
gas in the H{\sc ii} complex 
shows a filamentary distribution, which suggests the filamentary morphology of the parental molecular cloud. 

\section{Acknowledgements}
We are thankful to the anonymous referee for valuable comments. 
NP acknowledges financial support from the Department of Science \& Technology,
INDIA, through INSPIRE faculty award~IFA12-PH-36 at University of Delhi. The work at NCU is supported by the National Science Council through 
the grant no.~102-2119-M-008-001. This publication makes use of data from the Two Micron All Sky Survey (a 
joint project of the University of Massachusetts and the Infrared Processing 
and Analysis Center/ California Institute of Technology, funded by the 
National Aeronautics and Space Administration and the National Science 
Foundation), archival data obtained with the {\it Spitzer Space Telescope}
(operated by the Jet Propulsion Laboratory, California Institute 
of Technology, under contract with the NASA).
\bibliography{r}
\end{document}